\documentclass{iopjournal}
\usepackage{amsmath}
\usepackage{amssymb}
\usepackage[square,sort,comma,numbers]{natbib}

\usepackage{xcolor}

\usepackage{makecell}
\begin{document}

\articletype{Paper}

\title{The Gravitational-Wave Power Gap in Core-Collapse Supernovae: Insights from 60 Axisymmetric Simulations}

\author{
Haakon Andresen$^{1,*}$\orcid{0000-0001-7260-8983},
Xingzao Li$^{}$\orcid{},
Aurore Betranhandy$^{}$\orcid{},
Evan P. O'Connor$^{1}$\orcid{0000-0001-7368-8313},
Shuai Zha$^{2}$\orcid{0000-0001-6773-7830},
Sean M. Couch$^{3}$\orcid{0000-0002-6115-0156},
\\
}
\affil{$^1$The Oskar Klein Centre, Department of Astronomy, Stockholm University, SE-106 91 Stockholm, Sweden \\}
\affil{$^2$International Centre of Supernovae (ICESUN), Yunnan Key Laboratory of Supernova Research, Yunnan Observatories, Chinese Academy of Sciences (CAS), Kunming 650216, People's Republic of China\\}
\affil{$^3$Department of Physics and Astronomy, Michigan State University, East Lansing, MI 48824, USA \\}

\affil{$^*$Author to whom correspondence should be addressed.}

\email{haakon.andresen@astro.su.se}

\begin{abstract}
We analyse the gravitational-wave emission from 60 two-dimensional core-collapse supernova simulations. The models cover a range of progenitors and equations of state. We focus on the narrow frequency interval in the gravitational-wave spectrum where the emitted power is strongly suppressed (the power gap) and how its central frequency relates to the physical properties of the simulations.
We find that the power-gap frequency exhibits strong and systematic correlations with the properties of the inner core of the forming neutron star, for example the sound speed, suggesting that the gap encodes information about the behaviour of matter at extreme densities. We further examine how well several mechanisms proposed in the literature account for the presence and evolution of the gap in our simulations. Finally, we explore a scenario in which the gap arises from destructive interference between a narrow oscillation mode and a broadband background signal, demonstrating that such an interaction can produce a sharp minimum in the emitted gravitational-wave power.
\end{abstract}

\section{Introduction}
Core-collapse supernovae are expected to emit gravitational waves (GWs) that
carry detailed information about conditions in the stellar core as well as the
dynamics of the explosion. GWs offer direct insight
into the inner regions of core-collapse supernovae, whereas traditional
electromagnetic observations can only indirectly probe the explosion mechanism.
On the other hand, the GW signals from supernovae are weak. For a source 10 kpc
from Earth, we expect strains on the order of $10^{-21}$, making detections
outside the Milky Way challenging
\cite{Szczepanczyk_21,Szczepanczyk_23b,Szczepanczyk_24,Abac_25}.
{However, GWs are sensitive to the dynamics of the supernova core and
can propagate unhindered trough the stellar envelope. Therefore, it is expected that detecting the GWs from a galactic core-collapse supernova will allow us to extract detail information about the underlying physics, such as the explosion mechanism \cite{powell_24}, the presence of the standing accretion shock instability \cite{Lin_23,Veutro_25}, the angular momentum of the core \cite{Abdikamalov_14,Pastor-Marcos_24,Akhmetali_26}, or the properties of the protoneutron star (PNS) \cite{Bizouard_21,Casallas-Lagos_23,Bruel_23,Morales_25}. We refer the reader to Mezzacappa \&  Zanolin \cite{Mezzacappa_24} for a review.}

\begin{figure}
\centering
\includegraphics[width=1.\linewidth]{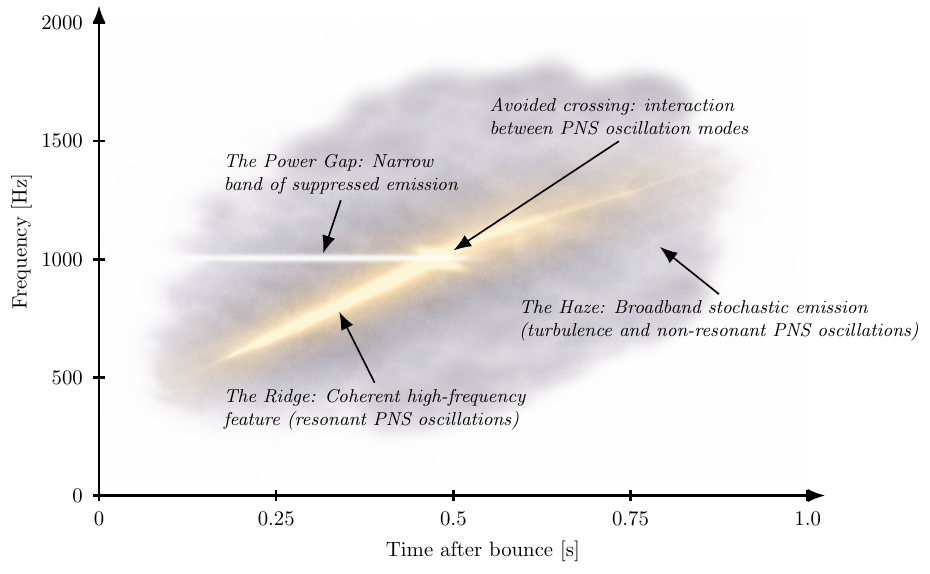}
\caption{Schematic of GW emission from core-collapse supernovae in the time–frequency domain. Yellow indicates the high-frequency ridge, the grey background represents the haze, and the narrow horizontal feature where emission is suppressed marks the power gap. }
\label{fig:overview}
\end{figure}

Multi-dimensional simulations predict broadband GW signals, with emission
between ${\sim}10\,\mathrm{Hz}$ and several kHz
\cite{Kotake_09,Murphy_09,Marek_09,Scheidegger_10,Yakunin_10,Kotake_11,
muller_e_12,muller_13,cerda-duran_13,Kuroda_14,Yakunin_15,Kuroda_16,
Andresen_17,Kuroda_17,Takiwaki_18,Hayama_18,Morozova_18,OConnor_18,
Radice_19,Andresen_19,Powell_19,Powell_20,Shibagaki_20,Mezzacappa_20,
Zha_20,Vartanyan_20,Andresen_21,Pan_21,Takiwaki_21,Eggenberger-Andersen_21,
Raynaud_22,Vartanyan_22,Jardine_22,Mezzacappa_23,Bugli_23,Vartanyan_23,
Powell_23,Jakobus_23,Pajkos_23,Richardson_22,Andresen_24,Richardson_24,Choi_24,
Murphy_25}.
While the emission spans a wide frequency range, most of the power lies between
$\sim100$ and $\sim1500\,\mathrm{Hz}$. Over the last decade, a coherent picture
of the GW emission from core-collapse supernovae has emerged from modern simulations.
Fig.~\ref{fig:overview} shows a schematic overview of the typical time-frequency structure 
of the high-frequency GW emission expected from core-collapse supernovae. Most simulations predict a narrow emission
component with a central frequency that increases with time (the bright golden feature in Fig.~\ref{fig:overview}), superimposed on a broad background (the grey diffuse cloud in the background of Fig.~\ref{fig:overview}).
In this work, we will refer to the narrow emission component as the \textit{ridge} and the background signal as the \textit{haze}. The narrow white gap seen in Fig.~\ref{fig:overview} depicts the \textit{power gap}, which is a recently discovered narrow band in which emission of GWs is strongly suppressed~\cite{Morozova_18}.
The frequency of the power gap typically remains stable in time, while the ridge frequency 
increases. The intersection of the power gap and the rising ridge tends to coincide with 
a change in the frequency evolution of the ridge. This change is likely due to an avoided 
crossing between two PNS oscillation modes, during which the character of the modes is 
exchanged. As a result, the mode associated with the GW emission changes character, and the evolutionary 
track of the ridge shifts accordingly (see, for example,~\cite{Vartanyan_23}).
Additional narrow-band
features can arise from various hydrodynamic phenomena. For example, from the standing accretion shock instability (SASI), which
introduces low-frequency emission between ${\sim}25$ and 250 Hz
\cite{Kuroda_16,Andresen_17}, or from oscillations in the central core
of the PNS, which may show decreasing central frequencies
\cite{Jakobus_23}.

The theoretical understanding of the processes responsible for GW emission has
been developed through direct analysis of numerical simulations and studies of
supernova oscillation modes
\cite{Kotake_09,Murphy_09,muller_13,Fuller_15,Sotani_16,Torres-Forne_18,
Morozova_18,Torres-Forne_19,Torres-Forne_19b,Sotani_19,Sotani_21,Andresen_21,
Rodriguez_23,Wolfe_23,Zha_24,Mori_23,Tseneklidou_25b,Sun_25}.  
However, the exact origin of the ridge has remained a
topic of debate for more than a decade.

Early works based on self-consistent two-dimensional simulations associated the
ridge with the deceleration of plumes impinging on the PNS surface or with
g-modes trapped in the PNS surface layer
\cite{Murphy_09,Marek_09,muller_13}.  
These studies also demonstrated that convection within the PNS can excite
surface g-modes and therefore contribute to the GW signal~\cite{muller_13}.
Murphy et al.~\cite{Murphy_09} showed that
plume deceleration in the PNS surface layer can reproduce 
both the observed frequencies and amplitudes of the GW signal.
With the advent of fully self-consistent three-dimensional simulations, the
relative importance of downflows versus PNS convection shifted in favour of PNS
convection as the dominant driver of the emission
\cite{Andresen_17,Mezzacappa_20}.  
By spatially decomposing the emission,Andresen et al~\cite{Andresen_17} found that most GW
power originates near the base of the PNS surface layer, where plumes from the
PNS convection zone overshoot into the stably stratified region.  
At the same time, Andresen et al.~\cite{Andresen_17} noted that the entire PNS contributes to
the high-frequency emission.  
Mezzacappa et al.~\cite{Mezzacappa_20} found that while PNS convection was the primary driver,
pressure modes within the convectively unstable layers dominated the emission, but that
surface g-modes contribute significantly as well.
However, other studies based on three-dimensional simulations 
emphasise the role of accretion.  
Radice et al.~\cite{Radice_19} argued that accretion onto the PNS is the main excitation
mechanism by demonstrating a strong correlation between the turbulent energy
accreted by the PNS and the total GW energy emitted.  
Similar conclusions were reached by several authors
\cite{OConnor_18,Powell_19,Morozova_18,Vartanyan_20}.  
Interestingly, Eggenberger Andersen et al.~\cite{Eggenberger-Andersen_21}, using a numerical setup similar
to the one employed here, found that both PNS convection and accretion-driven
downflows contribute significantly to the emission.

Mode analyses of the PNS have recently succeeded in matching the central
frequency of the ridge with eigenmodes of the PNS
\cite{Torres-Forne_18,Torres-Forne_19,Morozova_18,Sotani_16,Sotani_19,Westernacher-Schneider_19,Westernacher-Schneider_20,Sotani_21,Mori_23,Zha_24,Sotani_25},
shifting the interpretation away from localised perturbations toward coherent
global oscillations of the entire PNS.  
Such global oscillations involve larger coherent mass motions, which in principle should be more conducive to strong GW emission than small-scale turbulence.

The close agreement between the mode frequencies predicted by perturbative
analyses and those seen in simulations provides strong support for the
interpretation that the main GW emission arises from global oscillations of the
PNS. However, it is worth noting that mode analyses can only predict the frequencies of
oscillation modes that might exist. Perturbation analysis alone cannot predict if any given
mode is excited, how modes are excited, nor predict the strength of the GW emission associated with an oscillation mode.
Further support for the global-mode picture comes from the spatial agreement
between the quadrupole moments predicted by mode analyses and those extracted
directly from simulations~\cite{Eggenberger-Andersen_21,Zha_24,Murphy_25}.  
Recently, Murphy et al.~\cite{Murphy_25} emphasised that the region where a mode is driven
and the region contributing most strongly to the quadrupole moment need not be
co-located. The localisation of the excitation mechanism and the spatial origin of the GW signal
may therefore differ. 
{While these analyses have substantially advanced our understanding of the GW signal, they leave open how the emission is excited, how its energy is partitioned between the ridge and the haze, and what sets the power gap. Motivated by these open questions, in this work we use a suite of 60 two-dimensional simulations to examine the excitation of the GW emission and to study how the power gap varies with the underlying physical properties of the models.
We also explore various mechanisms for creating the gap and propose a new potential source of the power gap.}
% In this work, we use a suite of 60 two-dimensional simulations to examine the
% excitation of the GW emission and to study how the power gap varies with the underlying
% physical properties of the models. We assess how well existing explanations account for
% the power gap in our simulations. We also explore a new mechanism for creating the gap.

The paper is organised as follows. In section~\ref{sec:methods}, we describe our simulation framework, including the FLASH setup, the progenitor and EOS choices, and the extraction and processing of the GW signals. In section~\ref{sec:models}, we summarise the full set of 60 two-dimensional simulations employed in this study and outline their key properties. We investigate the mechanism responsible for GW emission in section~\ref{sec:gws}, separating the individual signal components. In section~\ref{sec:modes}, we present the results of our PNS mode analysis and 
discuss how the identified eigenmodes relate to the GW signals.
In section~\ref{sec:powergap}, we describe the correlation between the central frequency of the power gap and the physical properties of our simulations. We
review proposed explanations for the power gap and assess their compatibility with our simulations and investigate a new explanation for the power gap in section~\ref{sec:powergaporigin}. 
Finally, section~\ref{sec:conclusions} summarises our findings.

\section{Methods} \label{sec:methods}
\subsection{FLASH} \label{sec:flash}
The simulations were performed with a modified version of the \textsc{FLASH} framework (version~4)~\cite{Fryxell_2000}, specifically adapted for core-collapse supernova modelling~\cite{Dubey_09,Couch_13,Couch_14,OConnor_18,oconnor_18b}. Using adaptive mesh refinement, \textsc{FLASH} solves the Newtonian hydrodynamic equations with a modified general-relativistic effective potential (case~A in~\cite{Marek_06}). All models were evolved under the assumption of axisymmetry, employing 11 refinement levels with a finest grid resolution of 305 m.

Neutrino transport is handled using an energy-dependent M1 scheme
\cite{Cardall_12,Shibata_11} that evolves three neutrino species: electron neutrinos, electron antineutrinos, and a third species representing all heavy-lepton neutrinos and their antineutrinos. Neutrino opacities were generated with the \textsc{NuLib} library~\cite{OConnor_15}, using the
standard set of opacities outlined in~\cite{OConnor_18} with the addition of mean-field and virial corrections~\cite{horowitz_17}, and inelastic scattering on electrons~\cite{bruenn_85}.
Because \textsc{FLASH} groups all heavy-lepton neutrinos into a single species, pair-process reactions are treated approximately following~\cite{Burrows_06}. The effective emissivity is computed assuming isotropic emission, no final-state blocking, and integrating over the energy of the other neutrino pair. Effective absorption is then obtained via Kirchhoff’s law from the same emissivity. Further details regarding the opacity calculations can be found in~\cite{OConnor_15}.

\subsection{Progenitors}
The simulations are based on six progenitor models.  
Four of them correspond to solar metallicity with zero-age main-sequence masses of 15, 20, 23, and 28 solar masses ($M_{\odot}$)~\cite{woosley_07}, while the remaining two have zero metallicity and zero-age main-sequence masses of 35 and  85~$M_{\odot}$~\cite{Heger_10}.  
Progenitors with solar metallicity are labelled with the prefix \textit{s}, and those with zero metallicity with \textit{z}, followed by their respective initial masses. We will adopt a similar naming convention for the models based on the respective progenitors.

The six progenitors span a broad range in compactness, from 0.167 for \textit{s15} to 0.684 for \textit{z85}.  
The compactness parameter~\cite{OConnor_11} is evaluated at a mass coordinate of $M = 2.5\,M_{\odot}$ as,
\begin{equation}
\xi_{2.5} = \frac{2.5}{R(M_{\mathrm{bary}} = 2.5\,M_\odot) / 1000~\mathrm{km}}.
\end{equation}
For our set of progenitors, we find $\xi_{2.5}$ values of 0.167, 0.261, 0.428, 0.282, 0.482, and 0.684 for models \textit{s15}, \textit{s20}, \textit{s23}, \textit{s28}, \textit{z35}, and \textit{z85}, respectively. Our six progenitors span a broad range of compactness, which allows us to study the power gap and GW excitation across a range of accreation rates and PNS properties. 
{We selected progenitors spanning a large range in compactness to sample a variety of accretion histories and PNS properties. Using several progenitors rather than a single one allows us to study the power gap across a wide range of physical conditions.}

\subsection{Equations of State}
We employ ten finite-temperature nuclear equations of state (EOS): SFHo, SFHx~\cite{Steiner_13}, DD2~\cite{Typel_10}, and seven variations of the SRO EOS family~\cite{Schneider_17,Schneider_19}.
The SRO EOS~\cite{Schneider_19} is a
liquid-drop model with Skyrme-type interactions. 
The EOS model is characterised by the nuclear saturation density ($n_{\rm sat}$), the energy per baryon at saturation ($\epsilon_{\rm sat}$), the isoscalar incompressibility modulus ($K_{\rm sat}$), the symmetry energy ($\epsilon_{\rm sym}$) and its slope ($L_{\rm sym}$), the isovector incompressibility modulus ($K_{\rm sym}$), the effective nucleon mass at saturation density in symmetric nuclear matter ($m^*$), the neutron-proton effective-mass splitting at saturation density in pure neutron matter ($\Delta m^*$), and the pressures of symmetric and pure neutron matter at four times saturation density ($P^{(4)}_{\rm SNM}$ and $P^{(4)}_{\rm PNM}$).
A key feature of the SRO EOS is that it is distributed with an open-source code, allowing the user to produce custom EOS tables by varying the input parameters of the model. 
The baseline EOS adopts $n_{\rm sat}=0.155\,\mathrm{fm^{-3}}$, $\epsilon_{\rm sat}=-15.8\,\mathrm{MeV}$, $K_{\rm sat}=230\,\mathrm{MeV}$, $\epsilon_{\rm sym}=32\,\mathrm{MeV}$, $L_{\rm sym}=45\,\mathrm{MeV}$, $K_{\rm sym}=-100\,\mathrm{MeV}$, $m^*/m_n=0.75$, $\Delta m^*/m_n=0.10$, $P^{(4)}_{\rm SNM}=125\,\mathrm{MeV\,fm^{-3}}$, and $P^{(4)}_{\rm PNM}=200\,\mathrm{MeV\,fm^{-3}}$. 
We use seven different SRO EOS, varying one parameter at a time while keeping all others at their default values. 
In two of the SRO EOS variations we modify the effective nucleon mass at nuclear saturation density, with $m^*/m_n = 0.55$ and $0.95$, where $m_n$ is the free-neutron mass. The remaining five SRO EOS use symmetry–energy slope values of $L = 30,\, 38,\, 45,\,52,\, $ and $60\,\mathrm{MeV}$. 
This choice was motivated by our initial simulations, which showed noticeable differences between models computed with the SFHo and SFHx EOS. 
Since one of the distinctions between SFHo and SFHx is the value of the symmetry–energy slope they adopt~\cite{Steiner_13}, we wanted to examine whether the different choices of $L$ were responsible for the differences observed in the simulations.
{We selected our EOS to span a broad range of input physics, so that we can probe how the GW emission and the power gap respond to variations in the underlying microphysics. However, the EOS were chosen to be more or less consistent with current astrophysical and nuclear-experimental constraints \citep{Oertel_17,Tews_17}.}

\subsection{Gravitational Wave Extraction}
The GW signals are extracted from the simulations using the quadrupole formula. 
In the transverse-traceless gauge, the GW tensor
can be expressed in terms of two independent components, $h_{+}$ and $h_{\times}$, but $h_{\times}$ is zero in axisymmetric simulations. Far away from the source, at a distance $D$, in the slow-motion limit, and for an observer located
in the equatorial plane of the simulations, we can write the non-zero component as
\begin{align}\label{eqT:hp}
  h_{+} = \frac{3}{2}\frac{G}{c^4 D} \ddot{Q}_{33},
\end{align}
where $c$ is the speed of light and $G$ is
Newton's constant. $\ddot{Q}_{33}$ represents the second time derivative of the 
$zz$-component of the quadrupole moment and can be written as:
\begin{equation}
\label{eq:gws}
\ddot{Q}_{33} = \frac{4}{3} \frac{\mathrm{d}}{\mathrm{d}t} \bigg[  \int \, \rho  z v_z  \mathrm{d}^3 x \bigg],  
\end{equation}
where $z$ is the Cartesian coordinate along the $z$-axis, $v_z$ is the velocity component in the $z$-direction, and $\rho$ is the fluid density.
In this form, one of the time derivatives has been eliminated to avoid
numerical issues associated with second-order derivatives~\cite{oohara_97, finn_89, blanchet_90}. Eq.~\ref{eq:gws} is evaluated at every time step and the second time derivative is evaluated numerically using
\texttt{numpy.gradient}~\cite{numpy} during post processing.

We compute spectrograms by using short-time Fourier transforms (STFTs).
The STFTs are computed using \texttt{scipy.signal.stft} with a Blackman window~\cite{2020SciPy-NMeth}. Before plotting, we normalise the STFTs, such that their logarithmic values lie in the range $(-\infty, 0]$ and we use the same normalisation for every spectrogram. 
We filter the signals using high-pass and low-pass filters, removing any part
of the signals below 25 Hz and above 5000 Hz before calculating the STFTs.

\section{Model Description} \label{sec:models}
Here we give a summary of the 60 axisymmetric core-collapse simulations that form the basis for this work. The 60 simulations were first presented in~\cite{Li_24}, we refer the
reader to~\cite{Li_24} for a more detailed description.
Note that the power gap frequencies reported in this work differ somewhat
from those reported in~\cite{Li_24}, this is because we refined our
methodology for calculating the gap frequency for this work. Our method for detecting the power gap and for estimating the gap frequency is described in section~\ref{sec:powergap}.
For each of the six progenitors 
(\texttt{s15}, \texttt{s20}, \texttt{s23}, \texttt{s28}, \texttt{z35}, and \texttt{z85}) we performed 10 simulations using 10 different
nuclear equations of state (SFHx, SFHo, 55, 95, DD2, L30, L38, L45, L52, and L60). 
We label models by the progenitor name followed by the EOS (progenitor\_EOS).

\begin{figure}
    \centering
    \includegraphics[width=1.\linewidth]{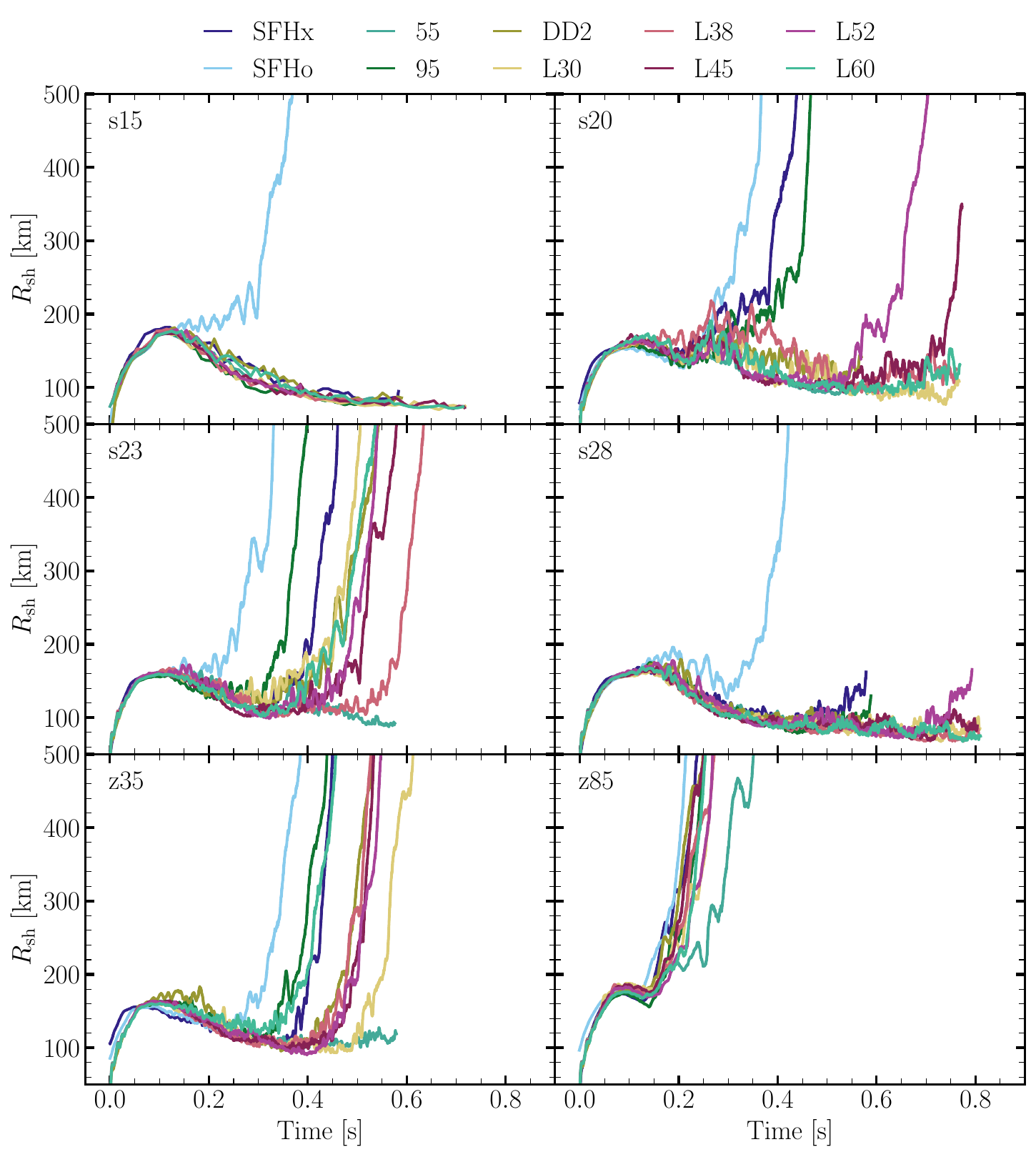}
    \caption{Average shock radius ($R_\mathrm{sh}$) as a function of time after bounce for all 60 models.
    Each panel shows the 10 EOS variations for a given progenitor, the progenitor name is indicated in the upper
    left corner. Each EOS is represented by one distinct line colour.}
    \label{fig:shock}
\end{figure}

At the end of each simulation, we find PNS radii mostly in the range 
$28$--$33$~km. A handful of models show larger PNS radii, $\sim 35$--$38$~km, these models are mostly based on the high compactness \texttt{z85} progenitor and some of the models are based on the \texttt{s23} progenitor.
The final PNS baryonic masses span $1.60-2.30\,M_\odot$ for the majority of the
simulations, but models based on the compact \texttt{z85} progenitor reach
$2.6-2.8\,M_\odot$. Central densities lie mostly in the $(5-7)\times 10^{14}\,\mathrm{g\,cm^{-3}}$ range for most models, but a few models that form black holes can reach values of $\sim 5\times 10^{15}\,\mathrm{g\,cm^{-3}}$. 
We note that the simulations were not evolved for the same duration and that many models 
were still evolving when the runs were terminated. It is therefore expected that the PNS 
properties of some models would continue to evolve, while others might yet explode or 
form black holes if evolved for longer times.
We show the average shock radius for all our models in Fig.~\ref{fig:shock}, where each 
panel corresponds to one progenitor, indicated in the top left corner. We define shock 
revival as the time when the average shock radius exceeds $300$~km. Several models 
experience shock revival, including \texttt{s15\_SFHo}, \texttt{s20\_SFHx}, 
\texttt{s20\_SFHo}, \texttt{s20\_95}, \texttt{s20\_L45}, \texttt{s20\_L52},
\texttt{s23\_SFHx}, \texttt{s23\_SFHo}, \texttt{s23\_95}, \texttt{s23\_DD2},
\texttt{s23\_L30}, \texttt{s23\_L38}, \texttt{s23\_L45}, \texttt{s23\_L52}, 
\texttt{s23\_L60}, \texttt{s28\_SFHo}, nine of the ten \texttt{z35} models 
(all except \texttt{z35\_55}), and all ten \texttt{z85} models.
Models \texttt{s28\_SFHx}, \texttt{s28\_95}, and \texttt{s28\_L52} do not reach the 
$300$~km threshold before the end of the respective simulations, but show a clear 
late-time expansion of the shock radius suggesting that shock revival is underway 
(see Fig.~\ref{fig:shock}).
By the end of our simulations, we found two black-hole-forming cases \texttt{z85\_SFHo} and \texttt{z85\_95}. 

A well-defined power gap is observed in most models. 
Out of the 60 simulations, 45 show a strong gap, 4 a weak one 
(\texttt{s23\_55}, \texttt{s23\_DD2}, \texttt{z35\_SFHx}, and \texttt{z85\_L45}), 
1 an unclear case (\texttt{s20\_95}), and 11 show no gap. 
The \texttt{s15}, \texttt{s20}, and \texttt{s28} progenitors exhibit strong and stable power gaps across all EOS, 
while the \texttt{s23} series shows mixed behaviour depending on the EOS, with \texttt{s23\_SFHo} displaying a clear gap and \texttt{s23\_SFHx} showing none. Interestingly, all models based on the \texttt{z85} progenitor lack a clear gap.

The key details for each of our models are listed in table~\ref{tab:model_summary}, the table lists the central power-gap frequency, the outcome of the simulations, as well as the time of shock revival for exploding models.

\begin{table*}[t]
\centering
\caption{Key properties for each of the 60 models presented in this work. The Power Gap column gives the central frequency of the power gap (see section~\ref{sec:powergap}). 
A dash indicates that a model does not have a well-defined power gap.
The Shock Revival column lists the time of shock revival (defined as when the average shock radius reaches $300$~km), measured in seconds after core bounce. Models that have not undergone shock revival by the end of the simulation are marked with a dash.
The \textit{BH} column indicates whether black-hole formation occurred during the simulated time. }
\begin{tabular}{lcccc@{\hskip 0.5cm}lcccc}
Model & \makecell{Power\\Gap\\{[Hz]}} & \makecell{Shock\\Revival\\{[s]}} & \makecell{End\\{[s]}} & BH & Model & \makecell{Power\\Gap\\{[Hz]}} & \makecell{Shock\\Revival\\{[s]}} & \makecell{End\\{[s]}} & BH \\ \hline \hline
\texttt{s15\_SFHx} & 1229 & --   & 0.58 & No & \texttt{s23\_SFHx} & --   & 0.42 & 0.54 & No \\
\texttt{s15\_SFHo} & 1200 & 0.31 & 0.52 & No & \texttt{s23\_SFHo} & 1307 & 0.28 & 0.50 & No \\
\texttt{s15\_55}   & 1127 & --   & 0.56 & No & \texttt{s23\_55}   & 1197 & --   & 0.57 & No \\
\texttt{s15\_95}   & 1159 & --   & 0.53 & No & \texttt{s23\_95}   & 1315 & 0.37 & 0.52 & No \\
\texttt{s15\_DD2}  & 1078 & --   & 0.59 & No & \texttt{s23\_DD2}  & 1144 & 0.48 & 0.55 & No \\
\texttt{s15\_L30}  & 1200 & --   & 0.72 & No & \texttt{s23\_L30}  & 1350 & 0.48 & 0.70 & No \\
\texttt{s15\_L38}  & 1200 & --   & 0.54 & No & \texttt{s23\_L38}  & 1300 & 0.60 & 0.65 & No \\
\texttt{s15\_L45}  & 1100 & --   & 0.72 & No & \texttt{s23\_L45}  & 1350 & 0.52 & 0.73 & No \\
\texttt{s15\_L52}  & 1100 & --   & 0.39 & No & \texttt{s23\_L52}  & 1300 & 0.51 & 0.71 & No \\
\texttt{s15\_L60}  & 1100 & --   & 0.71 & No & \texttt{s23\_L60}  & 1275 & 0.49 & 0.71 & No \\
\texttt{s20\_SFHx} & 1300 & 0.39 & 0.53 & No & \texttt{s28\_SFHx} & 1350 & --   & 0.58 & No \\
\texttt{s20\_SFHo} & 1270 & 0.32 & 0.48 & No & \texttt{s28\_SFHo} & 1248 & 0.39 & 0.54 & No \\
\texttt{s20\_55}   & 1172 & --   & 0.58 & No & \texttt{s28\_55}   & 1170 & --   & 0.59 & No \\
\texttt{s20\_95}   & 1216 & 0.45 & 0.53 & No & \texttt{s28\_95}   & 1210 & --   & 0.59 & No \\
\texttt{s20\_DD2}  & 1129 & --   & 0.57 & No & \texttt{s28\_DD2}  & 1171 & --   & 0.59 & No \\
\texttt{s20\_L30}  & 1250 & --   & 0.77 & No & \texttt{s28\_L30}  & 1200 & --   & 0.81 & No \\
\texttt{s20\_L38}  & 1250 & --   & 0.76 & No & \texttt{s28\_L38}  & 1200 & --   & 0.81 & No \\
\texttt{s20\_L45}  & 1225 & 0.76 & 0.77 & No & \texttt{s28\_L45}  & 1200 & --   & 0.80 & No \\
\texttt{s20\_L52}  & 1225 & 0.66 & 0.75 & No & \texttt{s28\_L52}  & 1150 & --   & 0.79 & No \\
\texttt{s20\_L60}  & 1200 & --   & 0.77 & No & \texttt{s28\_L60}  & 1150 & --   & 0.81 & No \\
\texttt{z35\_SFHx} & 1367 & 0.43 & 0.50 & No & \texttt{z85\_SFHx} & --   & 0.19 & 0.56 & No \\
\texttt{z35\_SFHo} & 1367 & 0.35 & 0.54 & No & \texttt{z85\_SFHo} & --   & 0.19 & 0.37 & Yes \\
\texttt{z35\_55}   & 1200 & --   & 0.58 & No & \texttt{z85\_55}   & --   & 0.29 & 0.50 & No \\
\texttt{z35\_95}   & 1240 & 0.40 & 0.53 & No & \texttt{z85\_95}   & --   & 0.22 & 0.44 & Yes \\
\texttt{z35\_DD2}  & 1157 & 0.50 & 0.56 & No & \texttt{z85\_DD2}  & --   & 0.20 & 0.48 & No \\
\texttt{z35\_L30}  & 1400 & 0.57 & 0.73 & No & \texttt{z85\_L30}  & --   & 0.22 & 0.62 & No \\
\texttt{z35\_L38}  & 1400 & 0.51 & 0.71 & No & \texttt{z85\_L38}  & --   & 0.22 & 0.61 & No \\
\texttt{z35\_L45}  & 1400 & 0.52 & 0.71 & No & \texttt{z85\_L45}  & --   & 0.21 & 0.61 & No \\
\texttt{z35\_L52}  & 1400 & 0.52 & 0.72 & No & \texttt{z85\_L52}  & --   & 0.22 & 0.62 & No \\
\texttt{z35\_L60}  & 1350 & 0.42 & 0.68 & No & \texttt{z85\_L60}  & --   & 0.23 & 0.62 & No \\
\end{tabular}

\label{tab:model_summary}
\end{table*}

\section{Gravitational Waves}\label{sec:gws}
\begin{figure}
\includegraphics[width=1.0\linewidth]{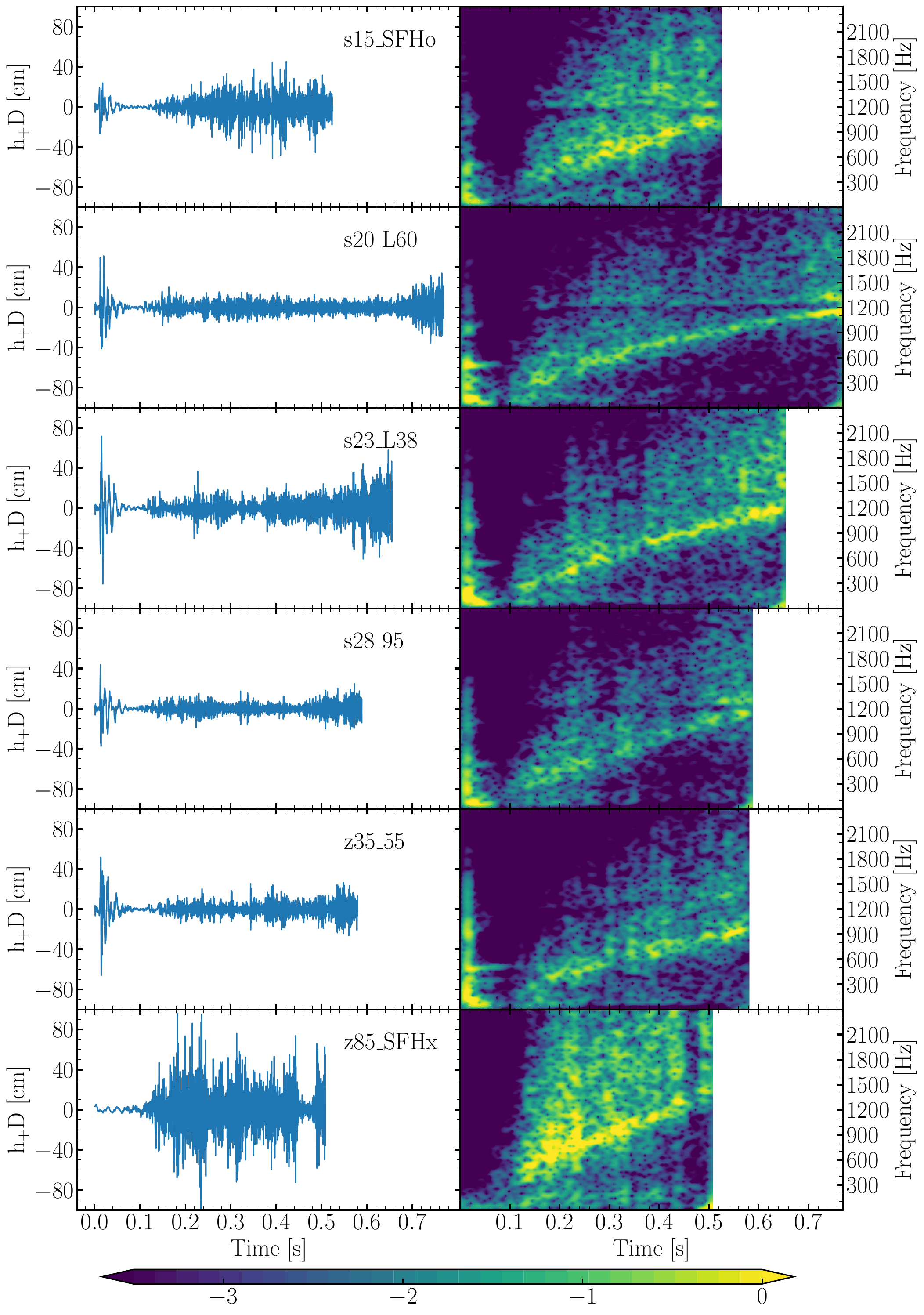}
    \caption{GW strains ($h_+$) multiplied by the distance to the observer ($D$) (left panels) and corresponding spectrograms (right panels) for a representative subset of our models. Each row shows one model, indicated in the top right of the waveform panel. Time is given in seconds after bounce. The colourbar is logarithmic and each spectrogram has been normalised by the same value, so that they can be directly compared.}
    \label{fig:signals}
\end{figure}
We show the GW amplitudes, strain multiplied by the distance to the event, and the corresponding spectrograms for a set of six representative models in Fig.~\ref{fig:signals}. Each row represents one model, with the waveforms shown in the left panels and the spectrograms in the right panels. The model names are indicated in the top right corners of the left panels. The spectrograms have been normalised by the same factor and can, therefore, be directly compared. All the models, except for \texttt{z85\_SFHx}, show an initial burst of emission 
associated with prompt convection directly after bounce. 
{Emission from prompt convection does not show up in model \texttt{z85\_SFHx}since it was started from a one-dimensional post-bounce profile. We did not evolve model \texttt{z85\_SFHx} through the initial collapse phase in 2D, rather we restarted it from a previous one-dimensional simulation. In the absence of rotation, the collapse phase is essentially spherically symmetric.}
The initial burst
is followed by a quiescent phase of roughly 0.1 s before the main emission
sets in once turbulence develops in and around the PNS. 
The signals typically have amplitudes of $\sim 20\,\mathrm{cm}$. We note, however, that 
axisymmetric simulations are known to overestimate the GW amplitude compared to 
full three-dimensional simulations, where the signal is typically a factor of 
$\sim 5$--$10$ smaller~\cite{Andresen_17}.
The exception is model \texttt{z85\_SFHx} which is based on a very compact progenitor and consequently emits a strong GW signal, reaching amplitudes exceeding $\sim 50\,\mathrm{cm}$. The models based on the z85 progenitor in general emit strong GW signals compared to models based on the other five progenitors. 

The signal morphology described in Fig.~\ref{fig:overview} is clearly visible in the 
spectrograms shown in the right column of Fig.~\ref{fig:signals}. The GW emission above 
250 Hz consists of two distinct components: a narrow emission band whose central frequency 
increases with time, forming a ridge in the spectrograms, and broadband background 
emission (commonly referred to as the haze~\cite{Vartanyan_20, Vartanyan_23}).
The background emission is particularly strong in model \texttt{z85\_SFHx}, where it is at times indistinguishable from the main emission ridge. A clear power gap is present in several models, but is absent in \texttt{z85\_SFHx} and weaker or difficult to identify in models \texttt{s23\_L38} and \texttt{z35\_55}.

At the end of the simulation, model \texttt{s20\_L60} exhibits a burst of
emission as the main emission mode approaches the power gap. This increase
in signal strength is accompanied by the appearance of an additional emission
feature just above the gap. The observed behaviour is likely indicative of an
avoided crossing between two modes, with the mode associated with the GW
emission undergoing a change in character. Models \texttt{s23\_L38} and \texttt{s15\_SFHo} show similar behaviour towards the end of their respective evolutions.

\subsection{Gravitational Wave Excitation} 
The main emission ridge is understood to be excited either by
downflows striking the PNS from above~\cite{OConnor_18,Radice_19,Powell_19,Morozova_18,Vartanyan_20} or from convection within the PNS~\cite{Andresen_17,Andresen_19,Mezzacappa_20,Andresen_21}. 
The exact nature of the haze has received less attention than the 
narrow ridge. Vartanyan et al.~\cite{Vartanyan_23} attributed the haze to accretion funnels striking the PNS, which is supported by the fact that the haze is seen to
diminish once the explosion develops and the accretion rate drops,
which is very clearly seen in the long simulations of~\cite{Vartanyan_23}.
It is possible that the haze is connected to non-resonant global
oscillations of the PNS~\cite{Zha_24}.
It is not obvious that the haze and the ridge have to be excited 
by the same mechanism. Furthermore, the relative contribution of the
haze and the ridge to the total GW energy is not clear.
The ridge appears as a narrow-band feature with relatively large instantaneous amplitudes. The ridge’s large amplitude suggests that it may contain significant power. On the other hand, it spans a limited frequency range, which could cause its total energy contribution, once integrated over frequency, to be modest.  
The haze, in contrast, exhibits smaller amplitudes but occupies a broad
frequency range and reaches higher frequencies. This extended bandwidth might 
allow the haze to carry a substantial amount of the total GW energy 
even though the signal is weaker at individual frequencies.

Exactly how much energy is carried by either component is important, because it influences how one should
interpret the correlations between the total energy emitted by
GWs and various properties of the simulations. We, therefore, 
start our analysis by decomposing the signal into two components, the haze and the ridge, and attempt to calculate the energy contained within each component separately. This is a delicate procedure, because there can be overlap between the two components. We, therefore, avoid reading too much into individual numbers and
instead focus on global trends seen in our large set of simulations. 
We also caution that our simulations are axisymmetric and that some of the conclusions we draw here might not carry over directly to three-dimensional simulations.

\begin{figure}
    \centering
\includegraphics[width=1.\linewidth]{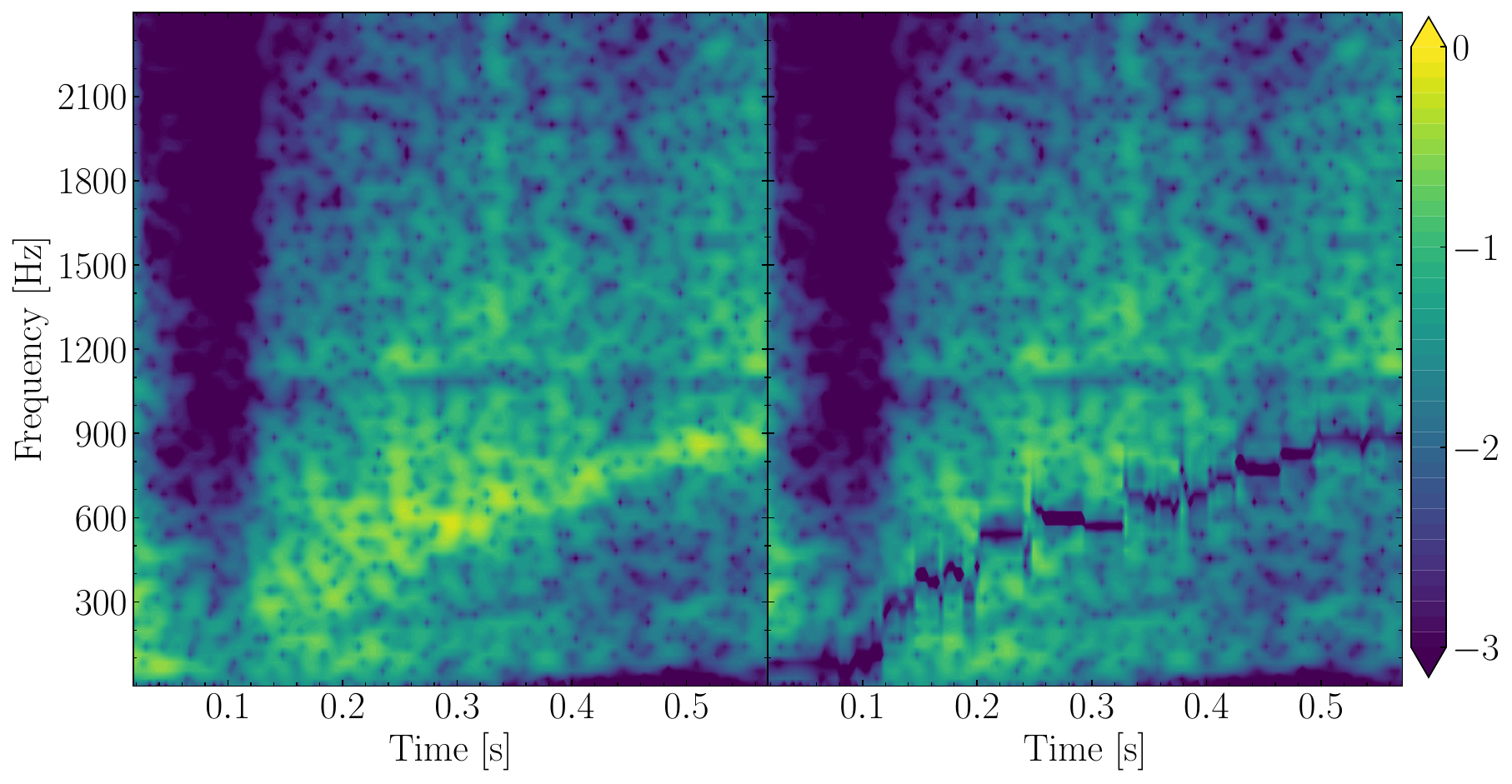}
    \caption{Example of the masking procedure described in section~\ref{sec:gws}. The figure shows the spectrograms for \texttt{s15\_DD2} with and without masking. The left panel shows the original spectrogram and the right panel shows the spectrogram with the main emission ridge masked out according to Eq.~\ref{eq:hazemask}. The spectrogram where the haze is masked is the complement of the spectrogram shown in the right panel. Time is given in seconds after bounce and the colour scale is logarithmic.}
    \label{fig:maskedspectrogram}
\end{figure}

To isolate the main emission ridge, we first compute spectrograms for each GW signal using a sequence of overlapping STFTs. 
Each STFT is evaluated with the \texttt{scipy.signal.stft} function.
This yields a time–frequency representation
\begin{equation}
Z(f,t) = \mathrm{STFT}\{h(t)\},
\end{equation}
where $|Z(f,t)|$ gives the instantaneous spectral amplitude.
We then determine the time-dependent peak frequency $f_c(t)$ of the signal by locating, for each STFT, the frequency at which 
$|Z(f,t)|$ is largest. In this way, $f_c$ tracks the dominant emission track.
After obtaining the peak-frequency as a function of time, we can
construct two masks that enable us to determine the energy contained in the two emission components:
\begin{align}
    w_{\mathrm{ridge}}(f,t) &= 
    \exp\Big[-\frac{1}{2}\,\frac{(f-f_c(t))^2}{\beta^2}\Big], \\
    w_{\mathrm{haze}}(f,t) &= 1 - w_{\mathrm{ridge}}(f,t)
    = 1 - \exp\big[-\frac{1}{2}\,\frac{(f-f_c(t))^2}{\beta^2}\big] \label{eq:hazemask},
\end{align}
where the width parameter $\beta$ controls how sharply the ridge is isolated. In the following, we will use $\beta = 100\,\mathrm{Hz}$.
The first mask $w_{\mathrm{ridge}}(f,t)$ selects the coherent power along the dominant emission track, while $w_{\mathrm{haze}}(f,t)$ captures the remaining background signal. 
{We note that any coherent emission that does not coincide with the ridge,
such as emission caused by weaker PNS modes or the SASI, 
is by construction included in the haze. The haze is best understood as anything but the the main ridge.}
These masks are subsequently applied to the power spectral density $|Z(f,t)|^2$ to compute the ridge, haze, and total energy contributions:
\begin{align} 
\label{eq:energiesr}
E_{\mathrm{ridge}}(t) &= \int w_{\mathrm{ridge}}(f,t)\, |Z(f, t)|^2\, f^2\, \mathrm{d}f, \\ 
\label{eq:energiesh}
E_{\mathrm{haze}}(t)  &= \int w_{\mathrm{haze}}(f,t)\, |Z(f, t)|^2\, f^2\, \mathrm{d}f, \\
\label{eq:energiest}
E_{\mathrm{tot}}(t)   &= \int |Z(f, t)|^2\, f^2\, \mathrm{d}f.
\end{align}
Note that the extra factor of $f^2$ comes from the fact that 
$E \sim \int \big((\partial_t h_+)^2 +(\partial_t h_\times)^2 \big) \mathrm{d}t$. Integrating Eqs.~\ref{eq:energiesr}, \ref{eq:energiesh}, and \ref{eq:energiest} over time gives the total
emitted energy for each component and the total energy.
Fig.~\ref{fig:maskedspectrogram} shows the masking procedure for
model \texttt{s15\_DD2}. 
We show only one representative example, since the procedure gives comparable
results across all models except \texttt{z85\_SFHx}.  
In the case of \texttt{z85\_SFHx}, the narrow emission ridge cannot be
clearly identified between 0.1 and 0.25 s post bounce.  
The reason is the presence of strong broadband emission during this period,
which makes it difficult to isolate the ridge, as seen in the bottom-right panel of
Fig.~\ref{fig:signals}.
Difficulties in distinguishing the ridge from the haze with our automatic
algorithm occur in a few other models as well, but the issue is most severe
for model \texttt{z85\_SFHx}.

\begin{figure}
    \centering
\includegraphics[width=1.0\linewidth]{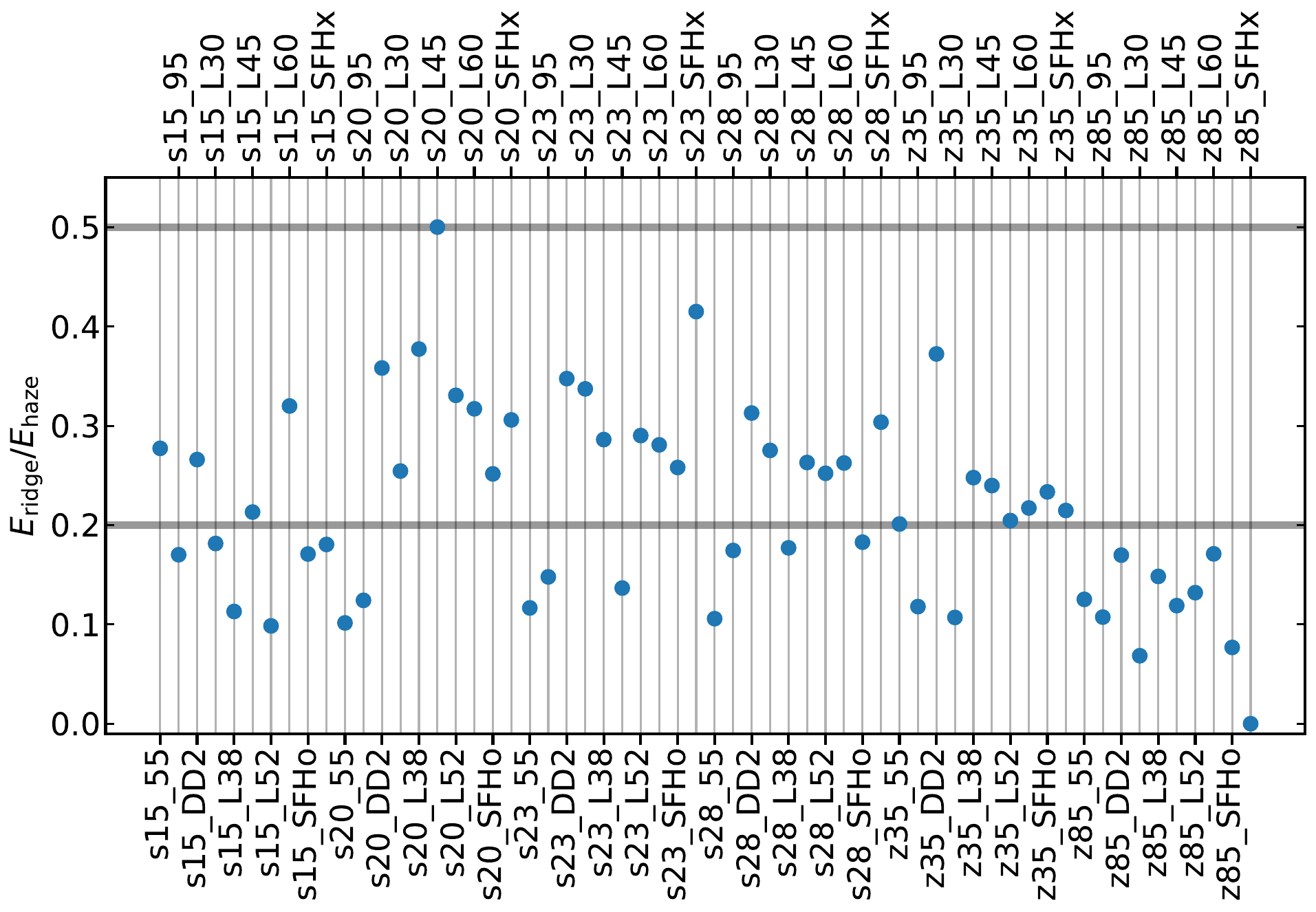}
    \caption{Ratio of the GW energy contained in the main ridge and the energy contained in the haze. Grey horizontal lines guide the eye to ratios of 0.2 and 0.5. Each dot corresponds to one simulation, where the model name is given on the x-axis above or below the plot.}
    \label{fig:ratio}
\end{figure}

In Fig.~\ref{fig:ratio}, we plot the ratio of the total time-integrated energy in the ridge to the energy in the haze for all our 60 models. Note that before we estimate the energy from our models, we apply bandpass filters to the GW signal and remove any part of the signal below 25 Hz or above 5000 Hz.
This plot has to be interpreted with caution for a few different reasons. Firstly, it is not trivial to extract $f_c$ as a function of time and stochastic bursts of emission can cause the
peak to shift away from the ridge for a handful of time windows. Secondly, the exact ratio depends on the choices we made when constructing our mask functions.
{The $\beta$--parameter sets the width of the ridge and, therefore, the exact ratios will depend on the choice of $\beta$. A different set of masking functions would also lead to small shifts in the power in the ridge and the haze.}
Thirdly (as mentioned above), in some models, for example \texttt{z85\_SFHx}, it can be difficult to distinguish the haze from the ridge. 
The haze can, in some cases, be so strong that there is no well-defined ridge.
Even with these caveats in mind, it is clear
that the haze contributes a significant part of the total emitted energy. In fact, the haze contributes the majority of the GW energy for all the models presented here.
This is in some sense not surprising, since the haze is a broadband emission feature and occupies a much larger area in the time-frequency domain than the ridge. Furthermore, the haze reaches significantly higher frequencies than the main ridge, which is important due to the $f^2$ scaling.

Radice et al.~\cite{Radice_19} showed that the total GW energy scales with the turbulent, kinetic plus thermal, energy accreted by the PNS. 
The accreted turbulent energy can be calculated by integrating the turbulent fluxes at the PNS surface over time~\cite{Radice_19}.
The turbulent fluxes are defined by first decomposing the fluid variables into mean and fluctuating components,
\begin{equation}
X = \langle X \rangle + X' ,
\end{equation}
where $\langle X \rangle$ denotes the angular average and $X'$ the deviation from it.
The corresponding angle-averaged rates of turbulent energy transfer are
\begin{equation}
    F_k' = \langle r^2 \left(\tfrac{1}{2}\rho v_i v_i\right) v_r' \rangle,
\end{equation}
and
\begin{equation}
    F_h' = \langle r^2 (\rho \epsilon + P) v_r' \rangle.
\end{equation}
Here $F_k'$ and $F_h'$ are the energy transfer rates corresponding to the turbulent kinetic and enthalpy flux densities, respectively. Additionally, $r$ is the spherical radius, $v$ is the velocity, $\epsilon$ is the specific internal energy, and $P$ is the pressure. 
A detailed derivation of these expressions is given in~\cite{Radice_16}.
The total accreted turbulent energy is then
\begin{equation} \label{eq:eturb}
    E_{turb}(r) = 4\pi\int \big(F_k'(r) + F_h'(r) \big) \mathrm{d}t,
\end{equation}
where $F_k'(r)$ and $F_h'(r)$ are evaluated at a given radius $r$.

To assess whether convective motions inside the PNS contribute significantly to the excitation of GWs in our models, we evaluate the turbulent energy not only at the PNS surface but also at the upper edge of the convectively unstable layer within the PNS. 
We define the PNS surface as the radius where the density falls 
below  $10^{11}\,\mathrm{g\,cm^{-3}}$.
Following~\cite{Andresen_17}, we use the turbulent mass flux
\begin{equation}
    f_m = \langle \rho' v_r' \rangle,
\end{equation}
to locate the upper boundary of the convective region.
The turbulent mass flux is negative within a convectively unstable layer. We define the bottom of the convective layer as the first point where the turbulent mass flux crosses zero and turns negative. Then, the top of the PNS convection zone is 
the radius where the turbulent mass flux becomes positive again.

\begin{figure}
    \centering
    \includegraphics[width=1\linewidth]{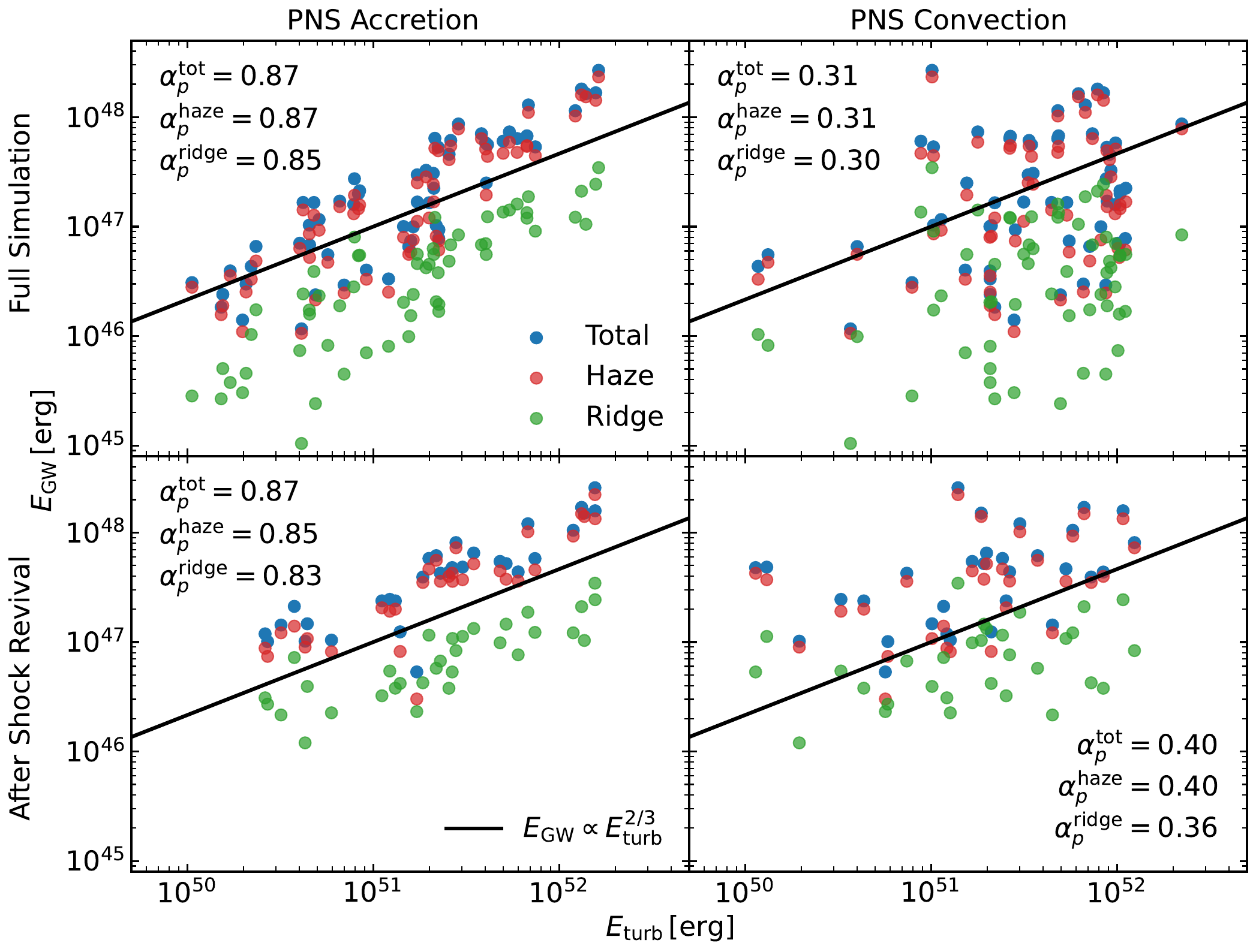}
    \caption{Correlation between the total GW energy and turbulent energy (Eq.~\ref{eq:eturb}) for all 60 models.
Blue points show the total emitted GW energy, red points the energy contained in the haze, and green points the energy in the ridge. Left column: correlation between the GW energies and the turbulent energy accreted onto the PNS from above. Right column: correlation between the GW energies and the turbulent energy passing through the upper boundary of the PNS convection zone. The top row shows correlations integrated over the entire simulation. The bottom row shows the same correlations, but restricted to the time after shock revival for the models that explode. {The Pearson correlation
coefficients ($\alpha_p$) between $E_{GW}$ and $E_{turb}$ (for the total signal, the ridge and the haze) are indicated in each panel.} }
\label{fig:correlation}
\end{figure}

The correlation between the energy emitted in GWs and the turbulent energy accreting onto the PNS and the turbulent energy passing through the upper boundary of the PNS convective region is shown in the left and right panels of Fig.~\ref{fig:correlation}, respectively. In both panels, we show the total energy emitted, the energy contained in the haze, and the energy in the ridge. We note that, for some models, the GW energy emitted
after shock revival accounts for nearly all the GW energy output.
These models develop early explosions and show substantial
GW emission following the onset of shock revival.
Consequently, they appear nearly indistinguishable in the two rows
of Fig.~\ref{fig:correlation}, as the emission after shock revival
dominates the total emitted GW energy.
{To quantify the strength of the linear trends, we compute the Pearson correlation
coefficient ($\alpha_p$) for $E_{GW}$ and $E_{turb}$, values close to $\pm 1$ indicating a strong linear correlation
and values near $0$ indicating no correlation. We compute $\alpha_p$ using the logarithmic values of $E_{GW}$ and $E_{trub}$.}
It is clear from Fig.~\ref{fig:correlation} that there is a strong correlation between the energy emitted as GWs and the turbulent energy accreted by the PNS from above.
{For the full signal duration, we find $\alpha_p = 0.87$, $0.87$, and $0.85$ for the total signal, the haze, and the ridge, respectively.}
Both the energy in the haze and the energy in the ridge correlate with the accreted turbulent energy. While higher turbulent energy accreted by the PNS tends to correlate with
more energy in the ridge component, the correlation shows some scatter.
On the other hand, for the haze we see a tighter and clearer correlation between the turbulent energy and the energy emitted by GWs.
{The turbulent energy at the top of the PNS convection layer shows a weak
correlation with the energy emitted in GWs, with $\alpha_p = 0.31$ for the
total signal, $\alpha_p = 0.31$ for the haze, and $\alpha_p = 0.30$ for the ridge
when evaluated over the whole signal duration, and $\alpha_p = 0.40$, $0.40$, and
$0.36$, respectively, after shock revival.}

Based on three-dimensional simulations, Murphy et al.~\cite{Murphy_25} showed that the dominant GW driving mechanism can evolve over time, with convection inside the PNS emerging as the main emission driver at late times.
The bottom row of Fig.~\ref{fig:correlation} shows the correlation between the turbulent energy fluxes inside and outside the PNS and the total GW energy 
for the subset of our models that undergo shock revival, except for models \texttt{s23\_L38} and \texttt{z85\_SFHx}. Shock revival sets
too close to the end of the simulation for the time integration to be meaningful for model \texttt{s23\_L38} and the ridge was difficult to separate from the haze for \texttt{z85\_SFHx}.
In this case, the time integration has been restricted to the time after shock revival.  We find no indication of a transition in the GW driving mechanism. 
The correlation between the accreted turbulent energy and the emitted GW energy remains strong, and no significant correlation is observed between the GW energy and the turbulent flux within the PNS convective layer.

The black lines in Fig.~\ref{fig:correlation} indicate a power-law scaling of $E_{GW} \propto E_{turb}^{2/3}$, which is close to the best-fit relations obtained when using both the full-duration signals and when only using the post–shock-revival emission for models that undergo shock revival. While there is significant scatter, a $2/3$ power law reasonably captures the overall scaling relation. The scaling is shallower than the quadratic dependence, $E_{GW} \propto E_{turb}^2$, reported by~\cite{Radice_19}. The relationship between $E_{turb}$ and $E_{GW}$ reported by Radice et al.~\cite{Radice_19} is physically well motivated. The GW energy radiated by non-radial oscillations is expected to scale quadratically with the energy of the oscillation mode responsible for the emission~\cite{osaki_73}, while the energy stored in such modes is, in turn, expected to scale linearly with the turbulent energy of the mass motions exciting them~\cite{goldreich_90,lecoanet_13}.
The fact that we find a shallower power-law scaling indicates that the transfer of energy from the turbulent forcing into GW emitting oscillation modes is not as efficient as the expected linear scaling. It is possible that the difference between
our findings and those of Radice et al.~\cite{Radice_19} is due to the fact that we
study two-dimensional simulations, while their results were based on full three-dimensional simulations. It is also possible that our larger dataset reveals a trend that is difficult to capture with a smaller set of numerical simulations. We found that a steeper power law
was preferred, for most of the cases, when fitting data points from individual EOS variations ($E_{GW}\sim E_{turb}^{0.8-1.6}$).
The shallower scaling we find could, therefore, reflect a combination of physical differences between 2D and 3D simulations and effects associated with sampling a broader range of models.

\section{Mode Analysis} \label{sec:modes}
{We compute the eigenmodes of the PNS using the framework presented in
\cite{Westernacher-Schneider_20} (see also~\cite{Zha_24}), restricting our
analysis to the dominant quadrupolar ($\ell = 2$) perturbations. Assuming
linear perturbations to a spherically symmetric background, we write the
Eulerian perturbations of the density ($\rho$), pressure ($P$), and
gravitational potential ($\Phi$) as
\begin{equation}
  \delta u(t, r, \theta)
    = \delta u(r)\, Y_\ell(\theta)\, e^{i \sigma t},
  \qquad u \in \{\rho, P, \Phi\},
\end{equation}
where $\sigma = 2\pi f$ is the angular frequency. The Eulerian fluid
displacement $\boldsymbol{\xi} = (\xi^r, \xi^\theta, 0)$ can then be written as
\begin{align}
  \xi^r       &= \eta_r(r)\, Y_\ell\, e^{i \sigma t}, \\
  \xi^\theta  &= \frac{\eta_\perp(r)}{r^2}\, \partial_\theta Y_\ell\,
                 e^{i \sigma t},
\end{align}
where $\eta_r$ and $\eta_\perp$ are the radial eigenfunctions corresponding
to the radial and tangential components of the displacement, respectively.
The linearised equations reduce to a first-order system
\begin{equation}
  \frac{d \mathbf{y}}{d r} = \mathbf{A}\, \mathbf{y},
  \label{eq:ode}
\end{equation}
for the state vector
\begin{equation}
  \mathbf{y} = (\eta_r, \eta_\perp, \delta\Phi, \partial_r \delta\Phi)^T ,
  \label{eq:state}
\end{equation}
where the coefficient matrix $\mathbf{A}$ depends on the background structure
through the sound speed, the adiabatic index, and the Schwarzschild
discriminant. The exact form of $A$ is given in Zha et al.~\cite{Zha_24}.
We solve this system by integrating outwards from a small radius near the
centre (one fifth of our finest grid spacing) out to a fixed outer radius
$R = 100\,\mathrm{km}$ \citep{Zha_24} where we impose the boundary condition
\begin{equation}
  \left[
    \partial_r\delta\Phi
    + \frac{\ell+1}{r}\,\delta\Phi
  \right]_{r=R} = 0.
\end{equation}
To isolate oscillations
of the PNS, we impose a vanishing Lagrangian pressure perturbation at the
PNS surface,
\begin{equation}
  \Delta P \big|_{r = R_{\mathrm{PNS}}} = 0 ,
\end{equation}
where, as above, we define the PNS surface as the radius at which the density
drops below $10^{11}\,\mathrm{g\,cm^{-3}}$. 
We have verified that our results do not qualitatively change
by shifting the definition of the PNS surface. We refer the reader to
\cite{Westernacher-Schneider_20} and \cite{Zha_24} for details regarding the mode analysis.
}

\begin{figure}
\centering
\includegraphics[width=1.\linewidth]{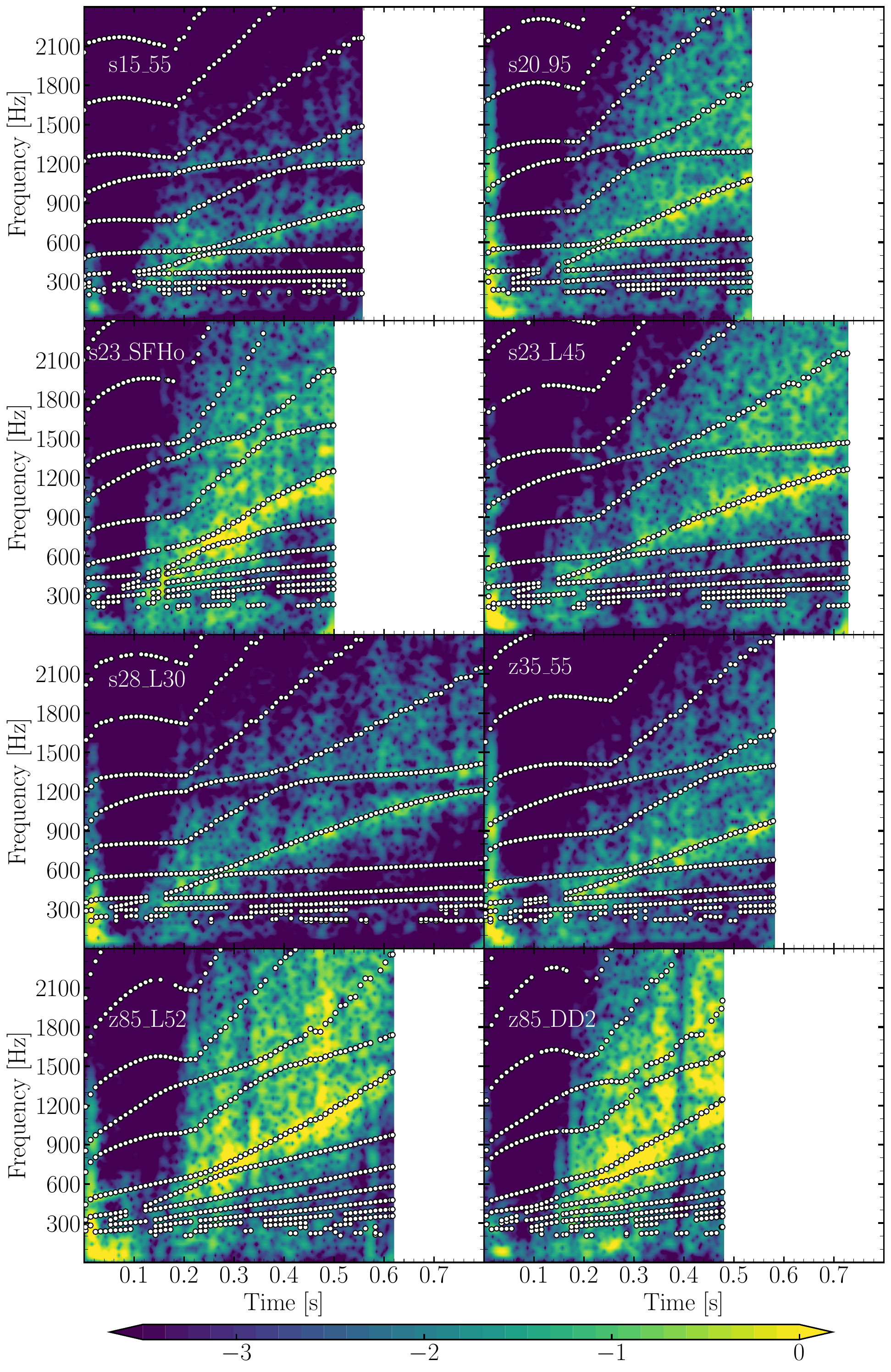}
\caption{Spectrograms of the GW for a few representative models.
Time is shown in seconds after bounce and the colour scale is logarithmic, the normalisation identical across all panels. White dots indicate the instantaneous eigenfrequencies of PNS oscillation modes obtained from the perturbative analysis.
The dots form tracks that trace the time evolution of individual modes. The model name is indicated in the top left corner of each panel. }
\label{fig:modetracks}
\end{figure}

\begin{figure}
    \centering
\includegraphics[width=1.\linewidth]{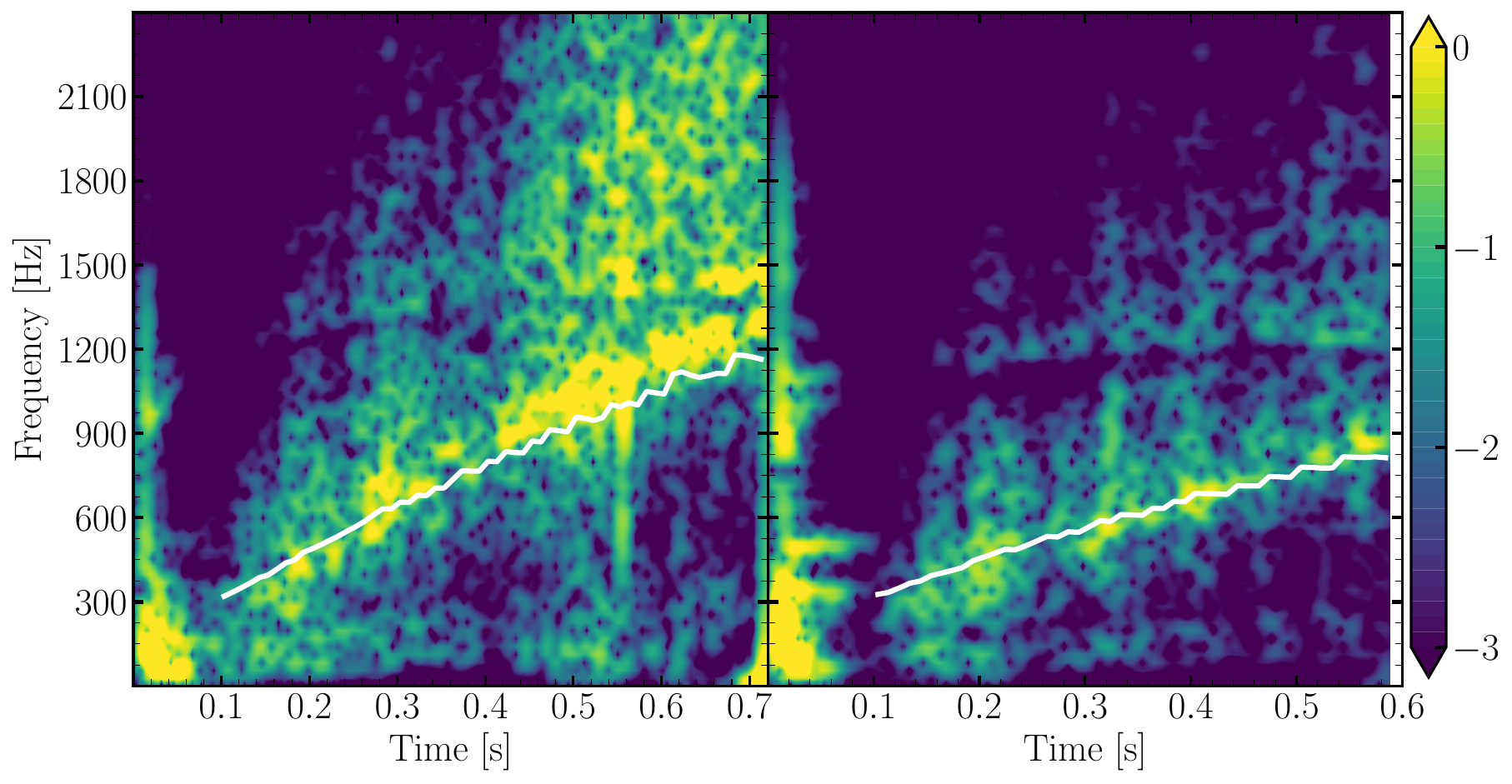}
    \caption{Spectrograms for models \texttt{z35\_L52} (left) and \texttt{s28\_55} (right).
    The white lines represent the inverse sound speed integrated from the centre of the PNS to the outer PNS radius, multiplied by a factor of 2.
    Time is given in seconds after bounce and the colour scale is logarithmic. }
    \label{fig:soundspec}
\end{figure}

We show the results of our perturbation analysis, for a handful of representative models, in Fig.~\ref{fig:modetracks}. The figure shows spectrograms with the instantaneous eigenfrequencies of PNS modes over-plotted as white dots. The dots form distinct tracks that trace the time evolution of the eigenmode frequencies. 
We find that the ridge is associated with a well-defined eigenmode in all of our models.
Since the classification of modes is dependent on the boundary condition we chose and the details of the analysis, we refrain from
attributing the GW emission unambiguously to either g-modes, p-modes, or the PNS f-mode.
However, we highlight an interesting and robust observation. 
The central frequency of the ridge is well approximated 
by the inverse of the sound-travel time across the PNS. 
In Fig.~\ref{fig:soundspec}, we show spectrograms for two models, 
\texttt{z35\_L52} (left panel) and \texttt{s28\_55} (right panel). 
The white curves in Fig.~\ref{fig:soundspec} show the inverse of the acoustic
crossing time of the PNS,
\begin{align} \label{eq:tcs}
t_{cs} = 2 \int_{0}^{r_{\mathrm{PNS}}} \frac{dr}{\langle c_s\rangle},
\end{align}
where $r_{\mathrm{PNS}}$ is the PNS radius and $\langle c_s\rangle$ is the angle-averaged
sound speed. The factor of two in Eq.~\ref{eq:tcs} accounts for the fact that a complete
oscillation involves an acoustic perturbation propagating from the centre to
the surface and then back again.  This round trip corresponds to one full
cycle. We see that $t_{cs}^{-1}$ closely traces the main emission ridge in Fig.~\ref{fig:soundspec}. In other words, the central GW frequency follows the acoustic crossing time of the PNS.
Consequently, the spectral properties of the ridge are compatible with sourcing by the PNS f-mode~\cite{Fuller_15,Sotani_19}. We emphasise, however, that similar correlations between the ridge central frequency and the expected g-mode frequency 
have also been reported in the literature~\cite{Murphy_09,muller_e_12,muller_13,Andresen_17}.

In addition to the mode that overlaps with the ridge,
we consistently identify a higher-frequency mode (we will call this mode 1) that lies immediately
above it in the spectrum.
For most of our models and for every model with a clear power-gap, this mode exhibits a nearly
flat frequency evolution and closely traces the upper boundary of the
power-gap region in the spectrogram for $t \gtrsim 0.3\,\mathrm{s}$
(we return to the correlation between this mode
and the power gap in section~\ref{sec:powergap}).
The next mode above the mode 1 branch
follows the power gap at earlier times ($t \lesssim 0.3$~s).
For most models, at around $t \sim 0.2-0.3$~s the two mode tracks approach each other
and subsequently separate without crossing.
After this interaction, the lower of the two branches
continues to trace the power gap.
The behaviour of the two modes is consistent with an avoided crossing
between neighbouring modes during the evolution of the PNS.
The behaviour is particularly visible for \texttt{s28\_L30},  \texttt{s15\_55}, and \texttt{s23\_L45}, 
see Fig.~\ref{fig:modetracks}.
A similar result was reported by~\cite{Morozova_18}, who found
that the frequency evolution of a mode identified as a core $g$-mode
traced the power-gap region in their models.

\section{The Power Gap} \label{sec:powergap}
To determine the power-gap frequency, we developed an automated detection routine 
based on short-time Welch spectra of the GW signal. 
The signal is divided into a series of overlapping time windows, and for each window we compute the power spectral density using 
\texttt{scipy.signal.welch} with a Blackman window. 
The resulting spectra are smoothed with a Savitzky–Golay filter (\texttt{scipy.signal.savgol\_filter}). 
We found that using 75 overlapping windows of 0.1~s width yields spectra in which the power gap is clearly identifiable. Prominent minima between 900 and 1500~Hz are then detected using \texttt{scipy.signal.find\_peaks}. We then average the location of the minima across all windows
and define a mean gap frequency.
Models based on the \texttt{z85} progenitor and model \texttt{s23\_SFHx} do not exhibit a well-defined power gap, according to both visual inspection of the spectrograms and the results of our automated detection algorithm.
\begin{figure}
    \centering
\includegraphics[width=1.0\linewidth]{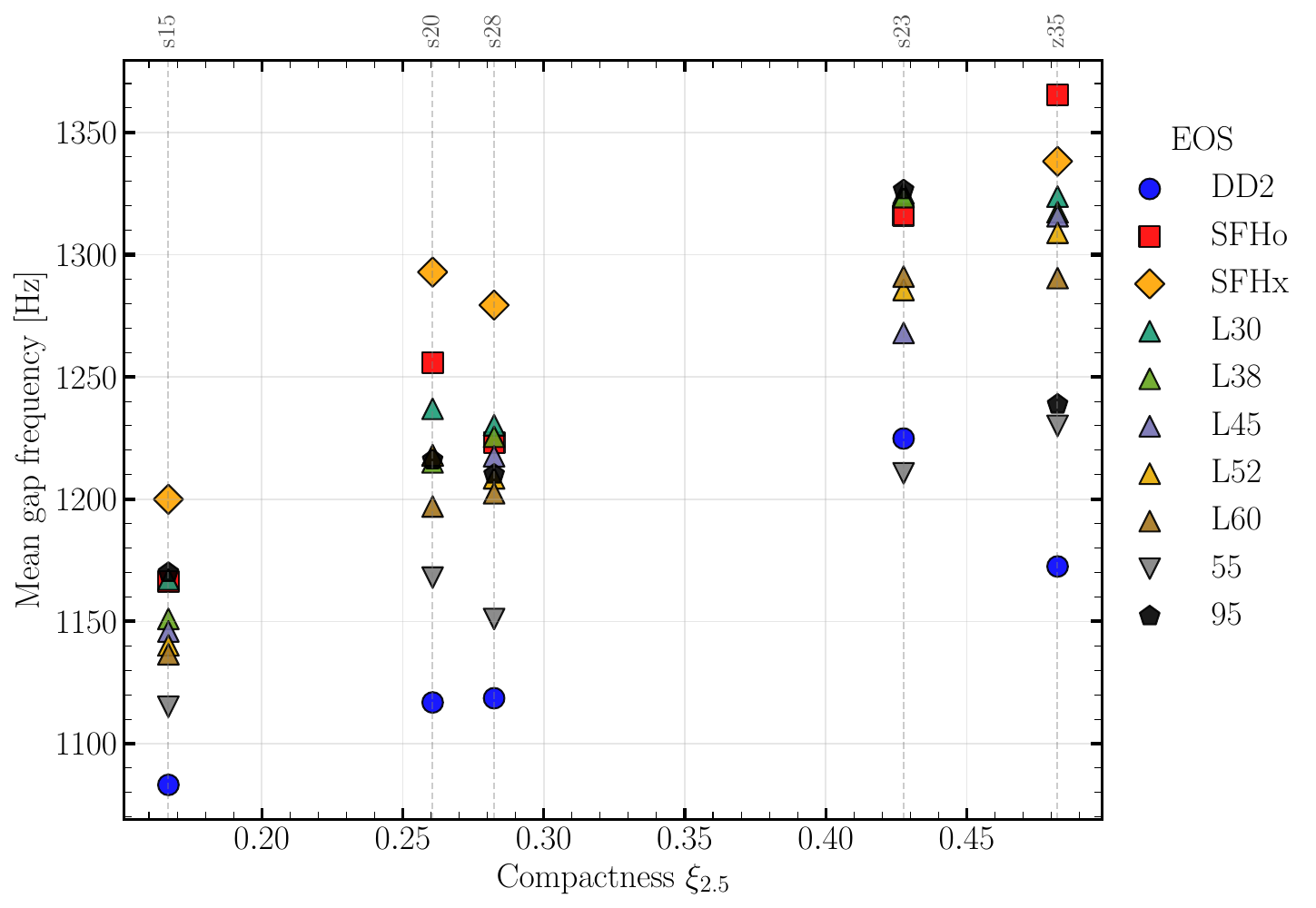}
    \caption{Mean power-gap frequency as a function of progenitor compactness, $\xi_{2.5}$.
Each marker represents an individual simulation, with colour and marker shape indicating
the EOS. Vertical dashed lines mark the progenitors on the compactness axis.}
    \label{fig:gap_c_eos}
\end{figure}

In Fig.~\ref{fig:gap_c_eos}, we show the central power-gap frequency for each model 
as a function of the progenitor compactness $\xi_{2.5}$, with different symbols 
indicating different EOS.
Two tentative trends can be identified. 
First, the average gap frequency increases with progenitor compactness: 
models with larger $\xi_{2.5}$ generally exhibit higher gap frequencies, 
whereas low-compactness progenitors show gaps at lower frequencies.
Second, at fixed progenitor compactness, the gap frequency varies systematically 
with the EOS.
Although there is no one-to-one mapping between EOS and bulk PNS properties, 
we find that EOS that produce more compact (i.e., more compressed) PNS tend to 
exhibit higher gap frequencies.
To quantify the degree of compression, we evaluate the PNS compactness 
$\xi_{\rm PNS} = GM/(Rc^2)$ and the central density $\rho_c$ at 
$t = 0.4$\,s post bounce for each model. 
Because progenitor structure also affects these quantities, we isolate the EOS 
dependence as follows: For each progenitor, we compute the percentage deviation 
of a given EOS model from the mean over all EOS for that progenitor. 
We then average these deviations over all progenitors (note that this effectively suppresses any outliers in the dataset). 
The resulting EOS-averaged deviations are shown in 
table~\ref{tab:eos_effect}.
The DD2 and SRO\_55 EOS produce PNS with lower compactness and significantly 
lower central densities compared to the average EOS. 
In contrast, SFHo, SFHx, and SRO\_95 yield systematically higher compactness 
and higher central densities, corresponding to more compressed PNS.
Consistently, simulations using SFHo and SFHx (red and yellow squares in 
Fig.~\ref{fig:gap_c_eos}) tend to occupy the upper part of the frequency range, 
whereas DD2 (blue circles) and SRO\_55 (grey inverted triangles) 
are predominantly found at lower gap frequencies. 
We emphasise, however, that these trends are not universal and are 
broken for several progenitors.
\begin{table}
\centering
\caption{PNS properties at $t = 0.4$\,s post bounce.
For each progenitor, we compute the percentage deviation of a given EOS model 
from the mean over all EOS for that same progenitor. The values listed here are 
the average of these deviations over all progenitors.
The PNS compactness is defined as 
$\xi_{\rm PNS} = GM/(Rc^2)$, evaluated at the PNS surface defined by 
$\rho = 10^{11}\,\mathrm{g\,cm^{-3}}$.
$\Delta X / \bar{X}$ denotes the mean relative deviation (in percent) from the 
progenitor-averaged value $\bar{X}$. }
\label{tab:eos_effect}
\begin{tabular}{lcc}
EOS & $\Delta \xi_{PNS}/\bar{\xi}_{PNS}$ [\%] & $\Delta\rho_c/\bar{\rho}_c$ [\%] \\
\hline\hline
\texttt{DD2}  & $-2.1$ & $-19.0$ \\
\texttt{SFHo} & $12.8$ & $15.4$ \\
\texttt{SFHx} & $0.8$ & $3.7$  \\
\texttt{55}   & $-7.7$ & $-13.4$ \\
\texttt{95}   & $6.2$ & $17.0$ \\
\texttt{L30}  & $-2.0$ & $1.3$  \\
\texttt{L38}  & $-2.0$ & $0.6$  \\
\texttt{L45}  & $-1.9$ & $-1.1$  \\
\texttt{L52}  & $-2.1$ & $-1.5$  \\
\texttt{L60}  & $-1.8$ & $-2.4$  \\
\end{tabular}
\end{table}
Across the $L$-series models the gap frequency decreases gradually with increasing $L$. However, the dependence of the power gap frequency on
the nuclear symmetry energy is weak and progenitor dependent. For example, the ordering is
broken for the simulations based on progenitor s23.

The weak trends we observe in Fig.~\ref{fig:gap_c_eos}
are likely a consequence of the non-linear interaction
of several aspects of the input physics, which together
determine the structure of the supernova core.
To better disentangle these effects, we therefore
investigate the direct correlation between selected
simulation properties and the power-gap frequency.
To quantify these correlations, we use the Spearman rank correlation
coefficient $\alpha_s$~\cite{spearman_1904}, as implemented in
\texttt{scipy.stats.spearmanr}.  
The Spearman rank coefficient ($\alpha_s$) measures the degree to which the relationship between two
variables can be described by a monotonic function.  
A value of $\alpha_s = 1$ $(\alpha_s = -1$) indicates a perfectly monotonic
increasing (decreasing) relationship.  
Values of $\alpha_s$ close to zero indicate that there is no strong evidence of a monotonic relationship, however
$\alpha_s = 0$ does not by itself provide evidence in favour of the null
hypothesis of no correlation.
\begin{figure*}
    \centering
    \includegraphics[width=0.49\linewidth]{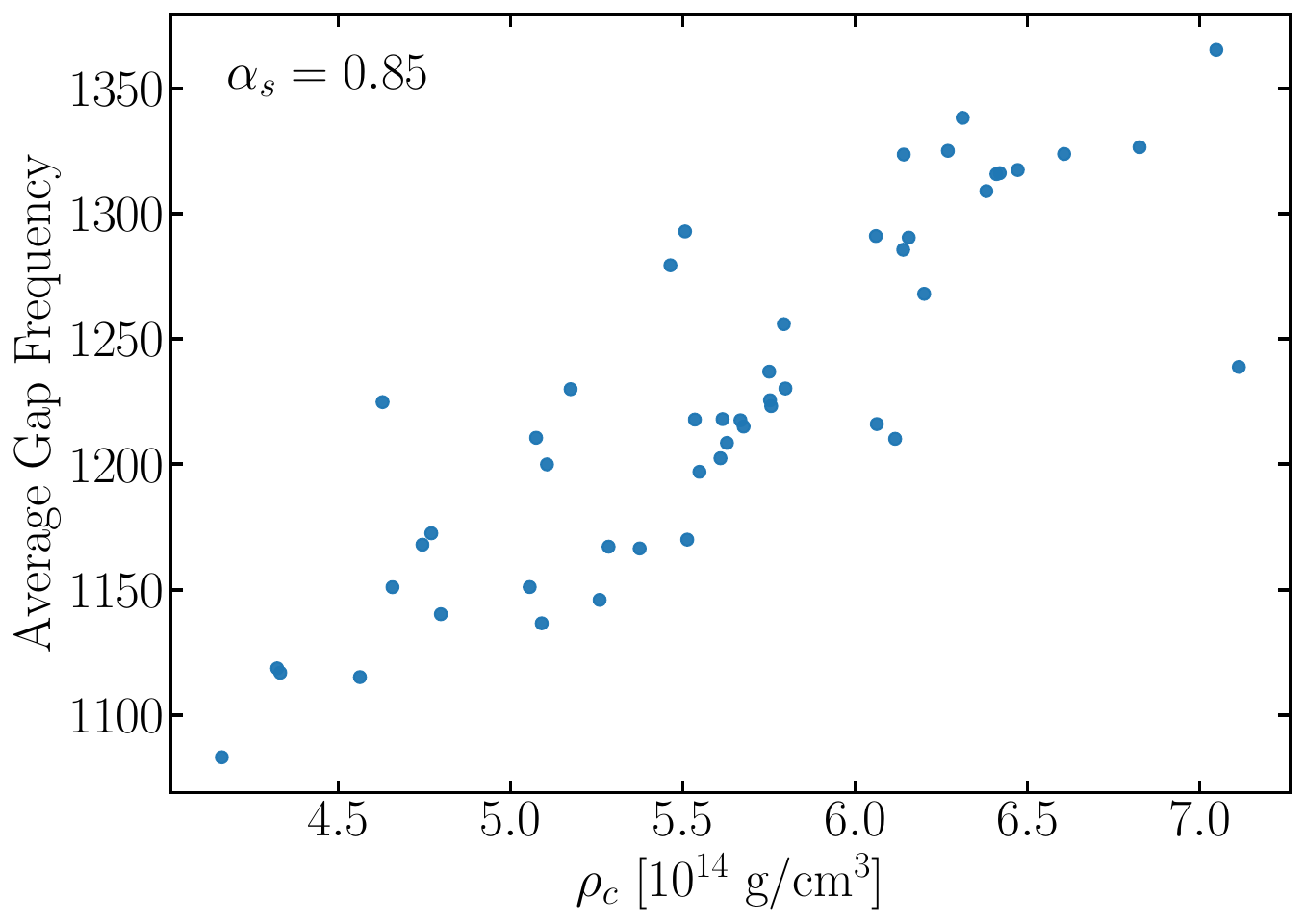}
    \includegraphics[width=0.49\linewidth]{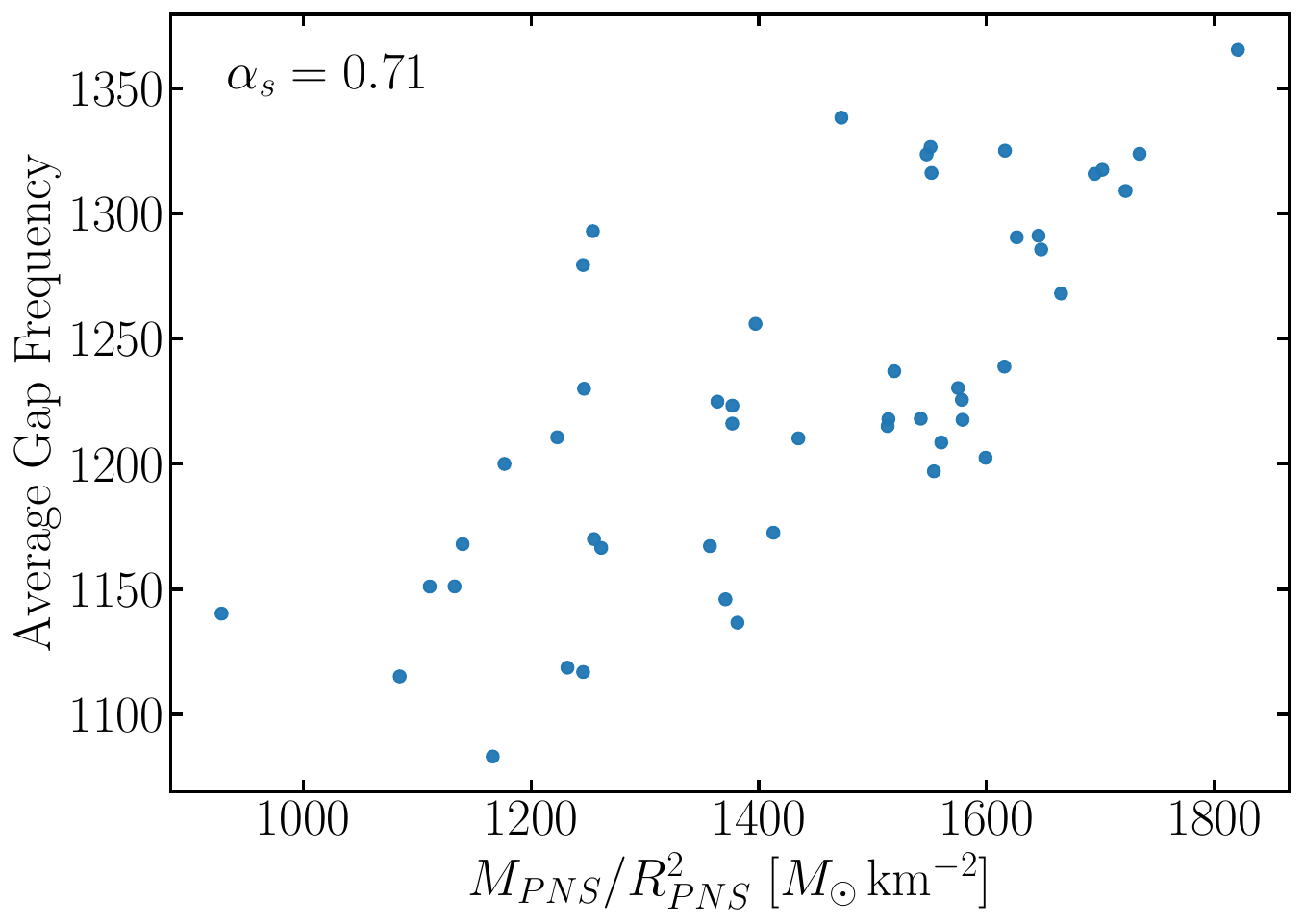} \\
    \includegraphics[width=0.49\linewidth]{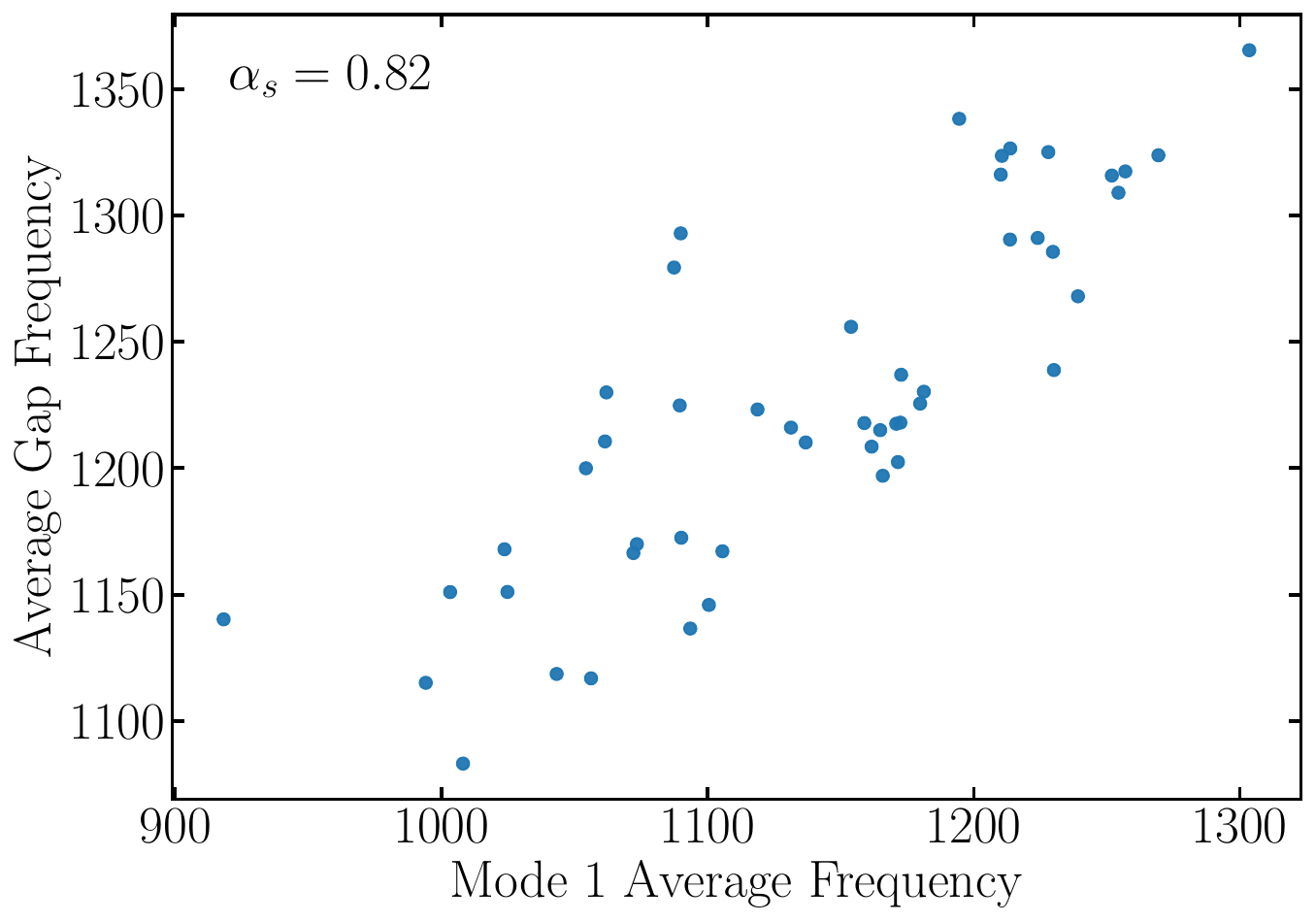}
    \includegraphics[width=0.49\linewidth]{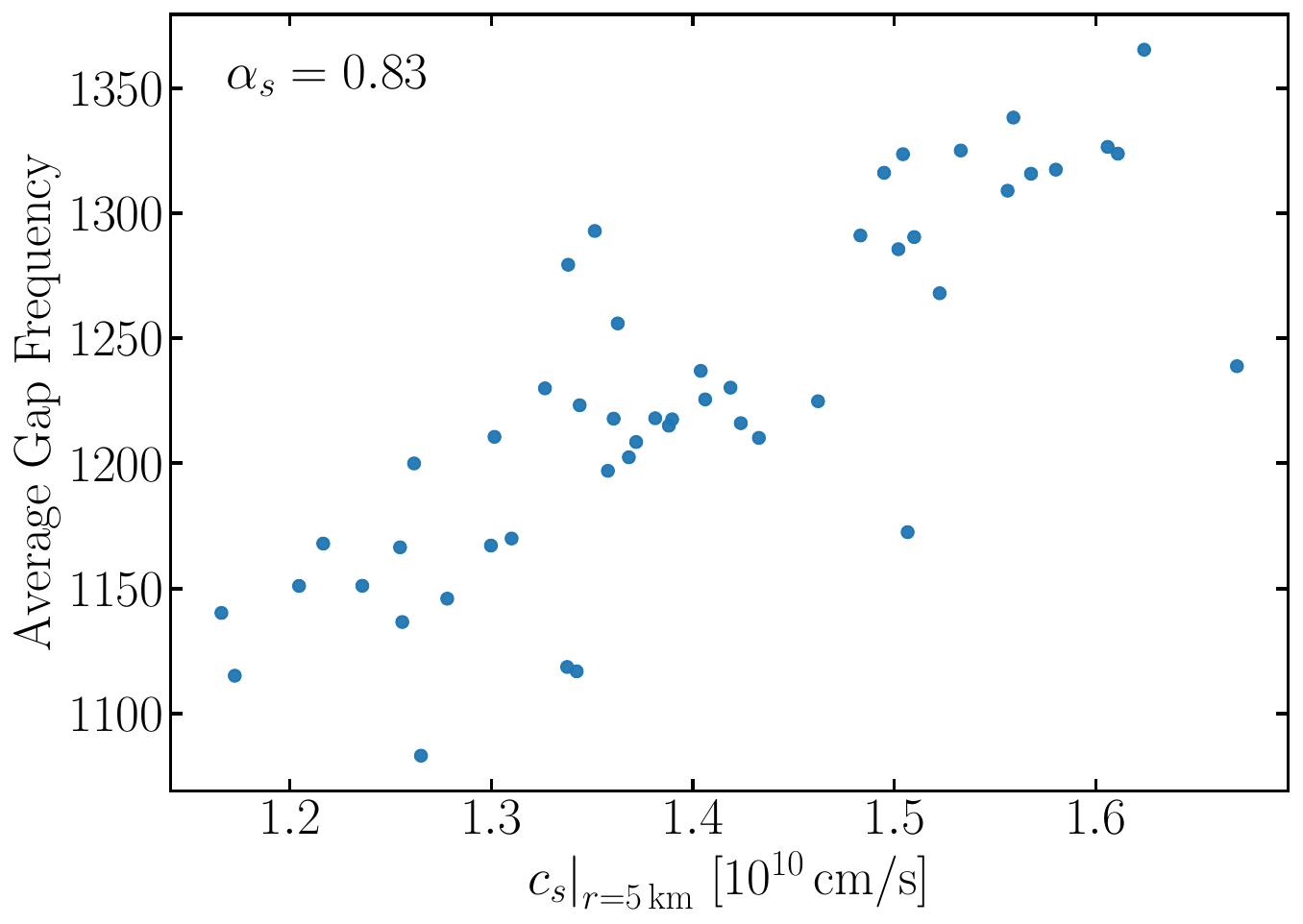} \\
    \includegraphics[width=0.49\linewidth]{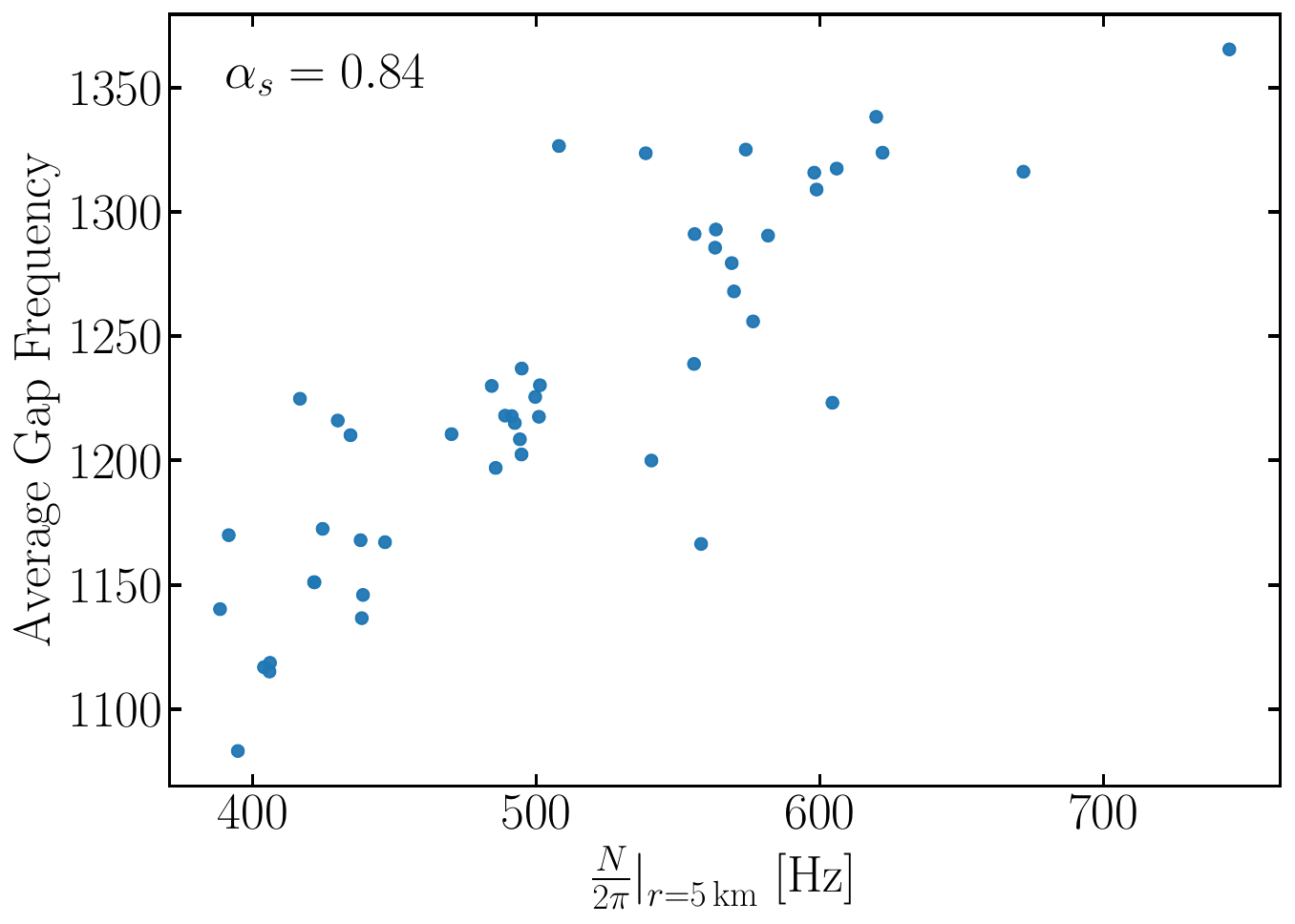}
    \includegraphics[width=0.49\linewidth]{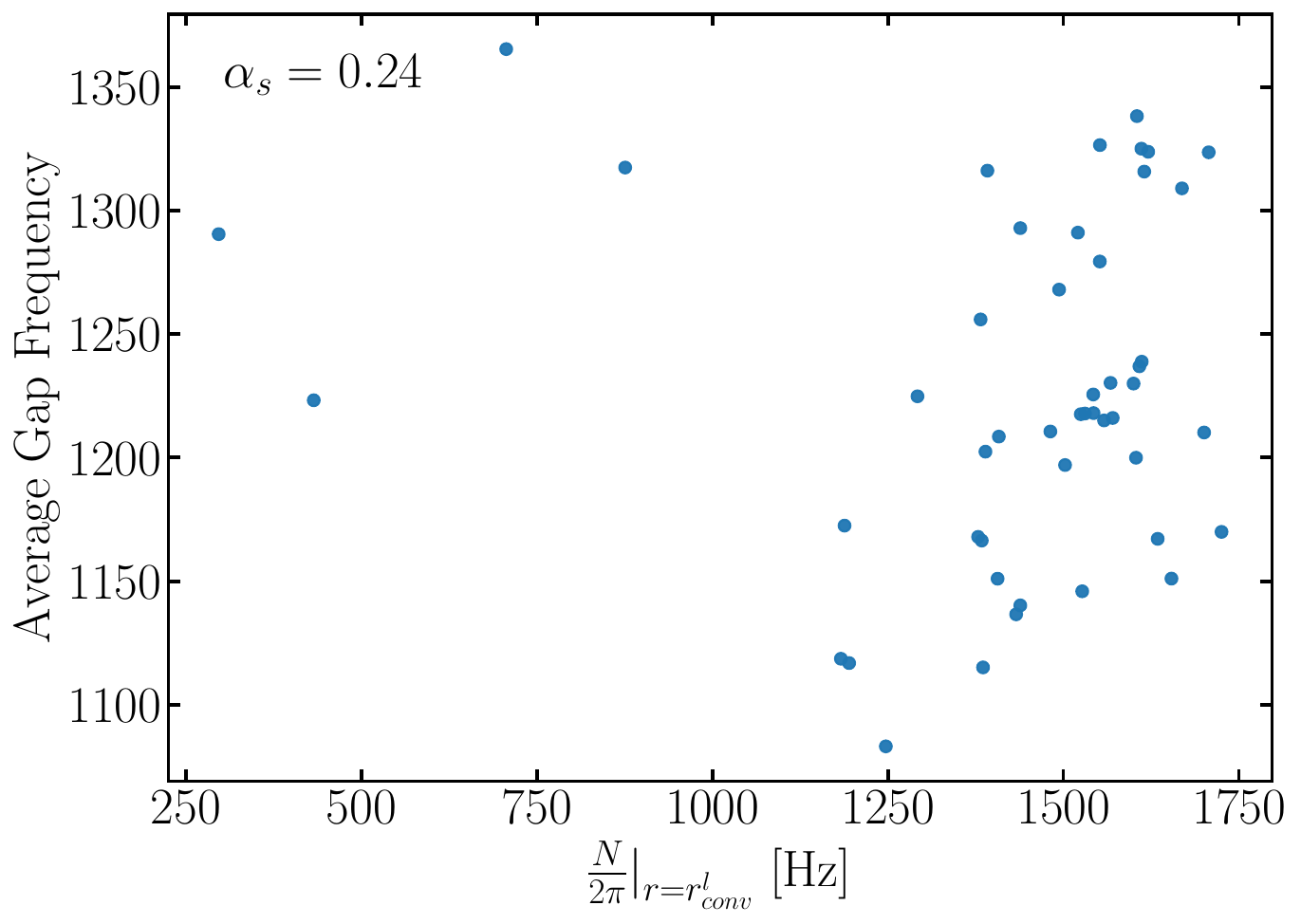}
    \caption{Correlation between the time averaged power-gap frequency and several time-averaged PNS properties for all models. Each blue dot represents one model.
    The different panels show scatter plots of the average gap frequency versus different PNS properties. Top left: central density $\rho_c$. Top right: PNS surface gravity. Middle left: the average frequency of PNS eigenmode~1 (see the text). Middle right: the sound speed at a radius of 5 km. Bottom left: the Brunt–Väisälä frequency at 5 km. Bottom right: the Brunt–Väisälä frequency evaluated at the outer boundary of the convective region. {In each panel, 
    $\alpha_s$ is the Spearman coefficient between the average gap frequency and the quantity shown in the panel.} }
    \label{fig:correlations}
\end{figure*}
We find clear correlations between the characteristic power-gap frequency and several key physical properties of the PNS.
The strongest correlation is obtained with the central density ($\alpha_s = 0.85$);
models with denser cores exhibit systematically higher gap frequencies.
This behaviour is consistent with the findings of Eggenberger Andersen et al.~\cite{Eggenberger-Andersen_21}, who reported that the power-gap frequency increases with central density across their EOS variations. 
The clear correlation between the power-gap frequency and the properties of the inner core is further reflected by strong correlations with the sound speed ($\alpha_s = 0.83$) and the Brunt–Väisälä frequency 
($\alpha_s = 0.84$), both extracted at a radius of 5 km.

The clear correlations between the power-gap frequency and both the sound
speed and the Brunt–Väisälä frequency in the core can be understood by recalling that
the Brunt–Väisälä frequency depends explicitly on the local sound speed:
\begin{equation} \label{eq:bvf}
N^{2} = \frac{1}{\rho}\,\frac{\partial \Phi}{\partial r}
\left[ \frac{1}{c_{s}^{2}}\,\frac{\partial P}{\partial r}
      - \frac{\partial \rho}{\partial r} \right].
\end{equation}
From Eq.~\ref{eq:bvf}, one might expect an inverse correlation between
the sound speed and the Brunt–Väisälä frequency, since $N^2$ contains an explicit
$1/c_s^2$ dependence. However, the situation is complicated by the pressure and density gradients.
Jakobus et al.~\cite{Jakobus_25} showed that the core Brunt–Väisälä frequency
may be estimated as
\begin{equation} \label{eq:ncore}
    N_{\rm core}^2 \sim c_{\rm gr}^2\, c_{\rm eos}^2
    {\frac{1}{r^2 \rho} \frac{\mathrm{d}s}{\mathrm{d}r}},
\end{equation}
where $\mathrm{d}s/\mathrm{d}r$ is the radial entropy gradient,
$c_{\rm gr}$ accounts for relativistic effects, and
\begin{equation} \label{eq:eosfac}
 c_{\rm eos} = \left(\frac{\partial P}{\partial s} \right)_{\rho,Y_e}^{1/2} c_s^{-1}.
\end{equation}
Here $Y_e$ denotes the electron fraction.
Although Eqs.~\ref{eq:ncore} and \ref{eq:eosfac} retain the explicit inverse
dependence on $c_s^2$, the effective EOS factor $c_{\rm eos}$ evolves slowly in time for two of the three models studied by~\cite{Jakobus_25},
while the Brunt–Väisälä frequency decreased significantly (see their Fig.~5).
Jakobus et al.~\cite{Jakobus_23} reported that the frequency of the
core $g$-modes in their models decreased as the sound speed at
twice nuclear saturation density increased.
In contrast, across our model set we find that models with
higher core sound speeds exhibit higher Brunt–Väisälä
frequencies in the core.

As discussed above, we identify two eigenmodes that trace
the power-gap region in many of our models (see Fig.~\ref{fig:modetracks}).
The time-averaged frequency of mode~1 (the mode immediately above
the ridge tracking mode in frequency) shows a strong correlation
with the time-averaged power-gap frequency ($\alpha_s = 0.82$), see Fig.~\ref{fig:correlations}. We compute the time-averaged frequency of mode~1 for
$t > 0.2$~s after bounce, as this branch traces the
power-gap region more clearly at later times.
We also tested whether averaging the frequencies of the
two modes that trace the power gap at different times
improves the correlation. Averaging both mode frequencies did not
lead to a significant change in the correlation strength. 

Interestingly, we do not find strong correlations when sampling the
sound speed and the Brunt–Väisälä frequency at the lower edge of the convective
layer. In this region we obtain $\alpha_s = 0.24$ for the correlation between the
power-gap frequency and the Brunt–Väisälä frequency, and $\alpha_s = 0.66$ for
the correlation with the sound speed.  
A coefficient of $\alpha_s = 0.66$ for the sound speed suggests a moderate
monotonic trend, whereas $\alpha_s = 0.24$ indicates no meaningful monotonic
relationship with the Brunt–Väisälä frequency in this part of the PNS.
This outcome is noteworthy because perturbation analysis has suggested that the
central frequency of the power gap may be connected to the Brunt–Väisälä
frequency at the base of the PNS convection layer~\cite{Zha_24}.

In Fig.~\ref{fig:correlations}, we also see a positive correlation between the power-gap frequency and the surface gravity, $M_{\mathrm{PNS}}/R_{\mathrm{PNS}}^2$, ($\alpha_s = 0.71$). We also found a weak correlation between the power-gap frequency and the size of the convective layer ($\alpha_s = 0.53$). 

\section{The Origin of the Power Gap} \label{sec:powergaporigin} 
Our results strongly indicate that the power gap is in some way
associated with the inner core of the PNS, which has been suggested by
several authors~\cite{Eggenberger-Andersen_21,Vartanyan_20,Morozova_18,Zha_24}. 
In the following section, we will review explanations of the power gap from the literature
and demonstrate yet another way the power gap could come about.

\subsection{Avoided Crossings} \label{subsec:crossing}
{Several studies have proposed that the GW power gap arises from an avoided
crossing between oscillation modes of the PNS
\cite{Morozova_18,Vartanyan_23,Wolfe_23,Bruel_23,Choi_24,Lella_26}. 
The basic picture suggested in these works is that two weakly coupled modes interact to suppress the 
emission at the frequency where they will eventually meet and undergo an avoided crossing.
It is not yet clear how this process functions or why it would occur at the avoided 
crossing frequency even when the frequencies of the modes are far apart.}
 
{To illustrate the dynamics of coupled modes and avoided crossings in this
context, we consider a simple analogy: two harmonic oscillators of masses $m_1$ and
$m_2$, with spring stiffnesses $k_1$ and $k_2$, respectively. The two oscillators are coupled by
a third spring with stiffness $k_c$.
Writing the displacements as $x_j(t) = A_j e^{i\omega t}$, the equations of
motion reduce to a standard eigenvalue problem whose two eigenfrequencies are
\begin{align}
  \omega_{\pm}^2
    = \frac{1}{2}\left(\omega_1^2 + \omega_2^2\right)
      \pm \sqrt{
        \frac{1}{4}\left(\omega_1^2 - \omega_2^2\right)^2
        + \frac{k_c^2}{m_1 m_2}
      },
  \label{eq:omega_general}
\end{align}
where $\omega_1^2 = (k_1 + k_c)/m_1$ and $\omega_2^2 = (k_2 + k_c)/m_2$ are the
natural frequencies of the two individual oscillators. For finite coupling
($k_c \neq 0$), the square-root term in Eq.~\ref{eq:omega_general} prevents
degeneracy and produces a finite minimum separation in $\omega^2$,
\begin{align}
  |\Delta(\omega^2)|
    = |\omega_+^2 - \omega_-^2|
    = \sqrt{
        \left(\omega_1^2 - \omega_2^2\right)^2
        + \frac{4 k_c^2}{m_1 m_2}
      },
\end{align}
which is minimised at the avoided crossing, where $\omega_1 = \omega_2$:
\begin{equation}
  |\Delta(\omega^2)_{\min}| = \frac{2 k_c}{\sqrt{m_1 m_2}}.
  \label{eq:gap_unequal}
\end{equation}
From Eq.~\ref{eq:gap_unequal}, we can make two interesting observations. Firstly, the mass normalised coupling between any two modes present in the PNS is relatively small.
We can draw this conclusion since the
minimum frequency separation between the modes observed in the GW spectrograms is small, see Fig.~\ref{fig:signals} or, for example, the spectrograms of~\cite{Vartanyan_23, Zha_24}.
Secondly, if we were to observe GWs from core-collapse supernovae we could work backwards and
extract the coupling strength by measuring the
mode separation width.}

The classical avoided crossing between the two normal modes can be demonstrated by keeping either $\omega_1$ or $\omega_2$ fixed and sweeping the other across the frequency spectrum. As the $\omega_+$ and $\omega_-$
approach each other, a deflection will occur as the two normal mode frequencies avoid crossing each other. 
Far from the crossing, each normal mode closely follows the trajectory of either $\omega_1$ or $\omega_2$.  
As $\omega_+$ and $\omega_-$ pass through the avoided-crossing region, they smoothly exchange character, with each mode continuing along the trajectory previously associated with the other. This exchange of character does not imply that the coupling between the modes is broken.  

Vartanyan et al.~\cite{Vartanyan_23} analysed long-duration three-dimensional simulations of core-collapse supernovae and explained the distinct ``bump'' observed in the GW frequency track around $\sim 1\,\mathrm{kHz}$ in their models as the result of a coupling between an inner $g$-mode and the outer $f$-mode of the PNS.  
In their picture, the coupling between the PNS surface and the inner core is gradually broken as the convective layer thickens, causing the normal mode to transition into a pure $f$-mode and produce the apparent frequency bump. In the simple coupled-mode picture considered here, however, such a feature can also arise naturally from continuous mode coupling, without requiring a physical decoupling of the cavities.

To understand why an avoided crossing between two coupled modes cannot by itself be responsible for the power gap, consider the
total Fourier power of the two modes
\begin{equation}\label{eq:modepower}
    P(\omega) = |F_+(\omega)|^2 + |F_-(\omega)|^2 + 2\,\Re(F_+(\omega)F_-^*(\omega)),
\end{equation}
where $F_\pm(\omega)$ are the Fourier amplitudes of the two individual modes.
The cross term can, in principle, produce partial destructive interference and reduce the power between the two modes, but the overlap between $F_+$ and $F_-$ is negligible far from the crossing. We, therefore, expect that the contribution from the two modes can only add
constructively. This can create a shallow valley of reduced power between the two peaks, but not a true spectral gap.  

In the simple case we have considered so far, in the absence of damping and forcing, the Fourier spectrum of the two oscillators is simply the sum of two delta functions
centred around $\omega_{\pm}$. This means that for oscillators without forcing or damping, the cross term vanishes even at the crossing.
Even if we allow the mode frequencies to evolve with time, the resulting spectrogram would only consist of two narrow ridges. In the spectrograms of core-collapse GWs we typically observe a background signal, superimposed on the main emission ridge.  Introducing a background into the simple model does not resolve this issue, since
adding an incoherent background component to the spectrum would merely fill in this valley. 
In the picture of two coupled modes and a background signal, the presence of a persistent and sharply defined gap in the GW spectrum requires destructive interference between a coherent background and one of the normal modes. Such anti-resonance, in turn, can only occur if at least one normal mode maintains an approximately fixed frequency in time. 
While this situation could arise, it is not per se 
related to the avoided crossing. This conclusion holds true even if
the mode power is spread out in a narrow frequency band 
(as could be the case in the presence of broadband forcing and/or damping), because the cross term in Eq.~\ref{eq:modepower} is small far away from the avoided crossing.

\subsection{A Zero in the Quadrupole Integral}
Zha et al.~\cite{Zha_24} performed perturbation analysis of the PNS for
perturbations with arbitrary frequencies.   
By relaxing the boundary condition necessary to obtain the eigenmodes of the PNS, they solved the perturbation equations for a range of
frequencies and obtained the corresponding radial profiles of the density perturbations, $\delta\rho(r,f)$, for quadrupolar ($\ell=2$) oscillations.  
The power emitted as GWs by a quadrupolar perturbation with radial profile 
$\delta\rho(r,f)$ is given by
\begin{equation} \label{eq:pertpower}
P(f) = C(f)\int_0^{R_{\mathrm{PNS}}} \delta\rho(r,f)\, r^2\, \mathrm{d}V,
\end{equation} 
where $C(f)$ is a normalisation factor~\cite{Zha_24}. Essentially, $C(f)$ 
encodes how effectively a perturbation at a frequency $f$ emits GWs.
Zha et al.~\cite{Zha_24} demonstrated that the perturbative power spectrum,
$P(f)$, exhibits a series of peaks and troughs across the frequency domain.  
The peaks in $P(f)$ correspond to frequencies at which the PNS can efficiently
emit GWs. In contrast, the minima arise at frequencies for which the radial structure of
the oscillation modes is such that the quadrupole is close to zero when integrated over the PNS, resulting in very weak GW emission at these frequencies.  

Zha et al.~\cite{Zha_24} found a pronounced minimum in $P(f)$  between 1.0 and 1.3 kHz, coinciding with the power gap in their simulations. Importantly, this minimum
remained stable in time~\cite{Zha_24}.  
Zha et al.~\cite{Zha_24} tentatively linked the minimum associated with the
power gap to the factor $1 - \omega^{2}/N^{2}$ appearing in the perturbation
equations, as before $\omega$ is the angular frequency of the perturbation.  
They found that the gap occurred at the frequency for which this term becomes
zero at the bottom of the PNS convective layer. In other words, when
$\omega$ equals the Brunt–Väisälä frequency at the bottom of the convective region.  
In addition, Zha et al.~\cite{Zha_24} found close agreement between the
spatial distribution of GW power predicted by the perturbative analysis and the
quadrupole moments calculated directly from their simulations, providing
further support for the perturbative interpretation. 
However, in our simulations we do not find evidence for a strong correlation
between the power-gap frequency and the Brunt–Väisälä frequency at the bottom
of the convective layer (see Fig.~\ref{fig:correlations}), which is in contrast to the trend suggested by
\cite{Zha_24}.

From the viewpoint of coupled modes superimposed on a background haze, the mechanism proposed by~\cite{Zha_24}
represents a zero in the background rather than an interaction
between the modes. For the models presented in~\cite{Zha_24}, the perturbation analysis revealed multiple frequencies at which the total GW power vanished when integrated over the PNS (see Fig.~6 of~\cite{Zha_24}). Most of these minima evolved in time, but one remained nearly stationary and coincided with the observed power gap. Zha et al.~\cite{Zha_24} therefore concluded that the emergence of the power gap depends on the stability of minima in the GW power emitted by the PNS as a function of frequency. Given that the central power-gap frequency can evolve with time~\cite{Vartanyan_23}, a slow secular drift of such a minimum is likely sufficient to produce a gap, rather than requiring strict stationarity.

\begin{figure}
    \centering
    \includegraphics[width=0.49\linewidth]{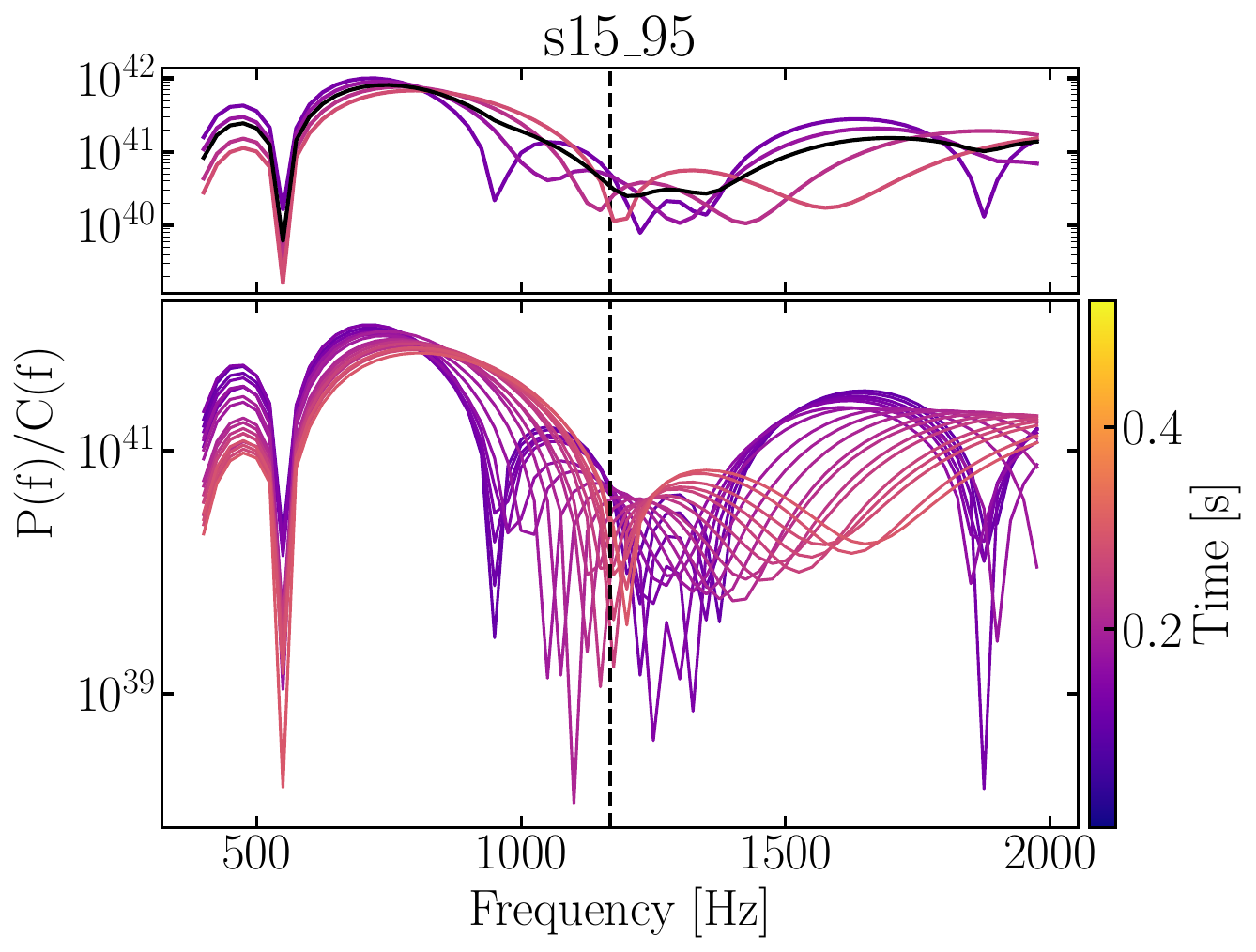}
    \includegraphics[width=0.49\linewidth]{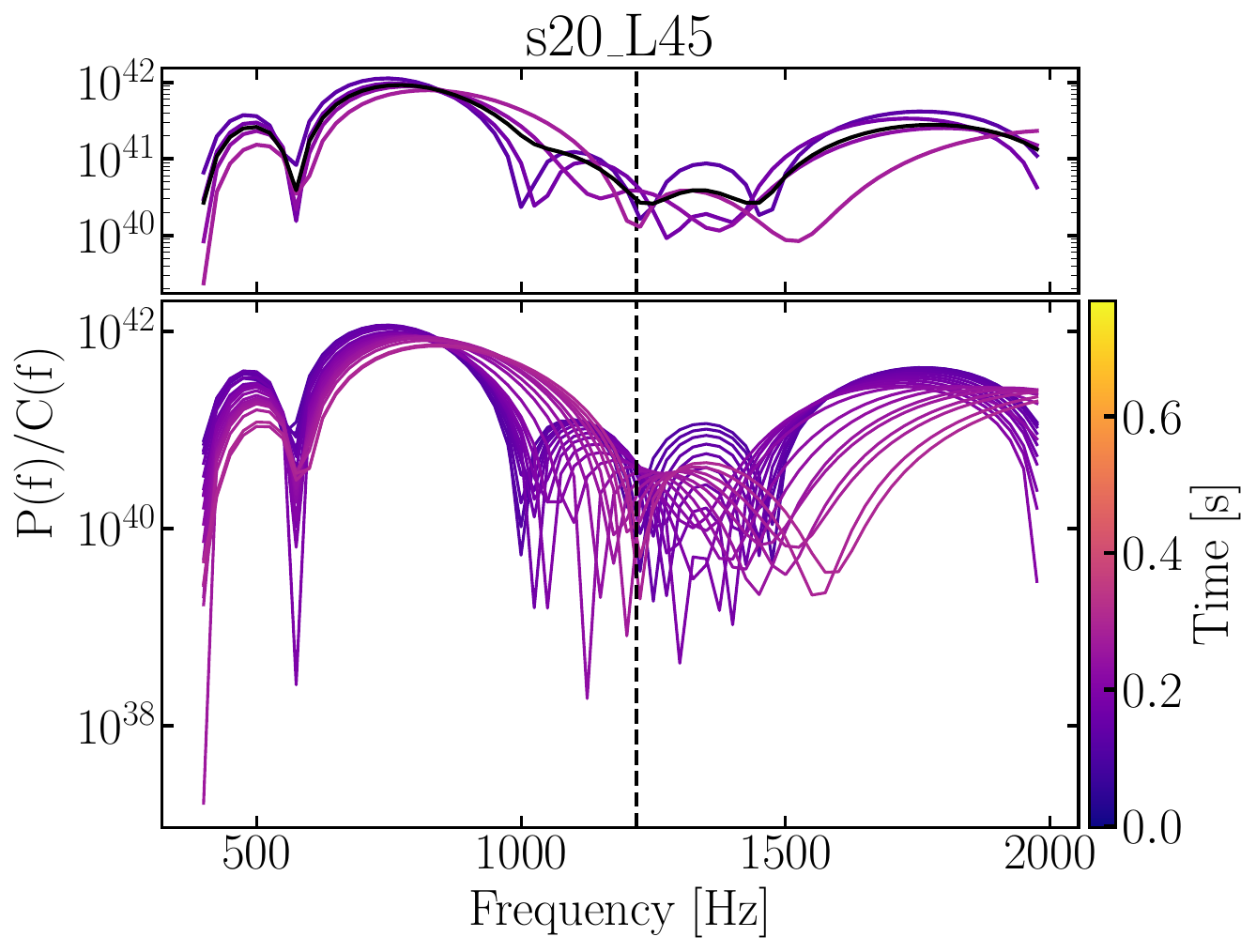} \\
        \includegraphics[width=0.49\linewidth]{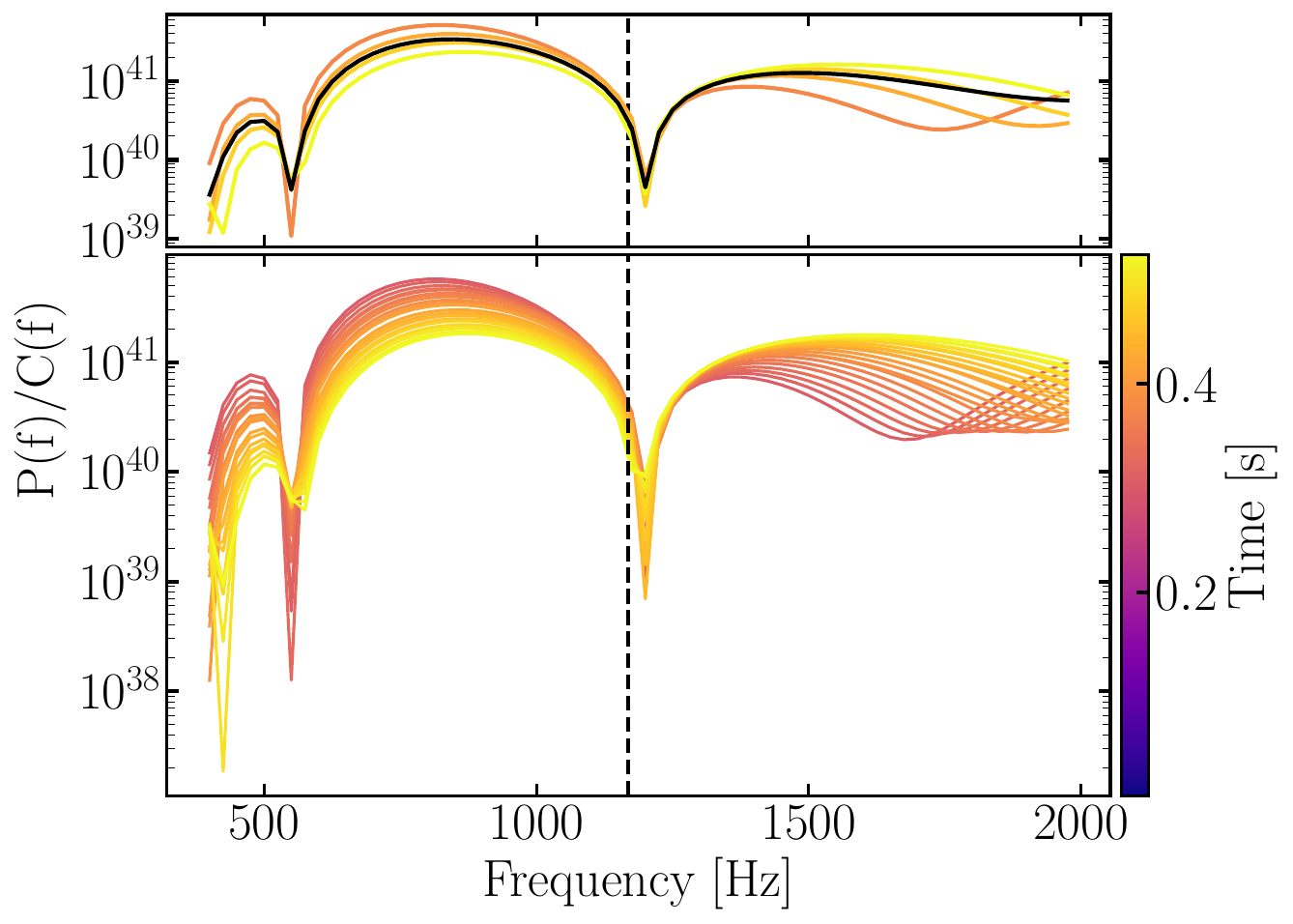}
            \includegraphics[width=0.5\linewidth]{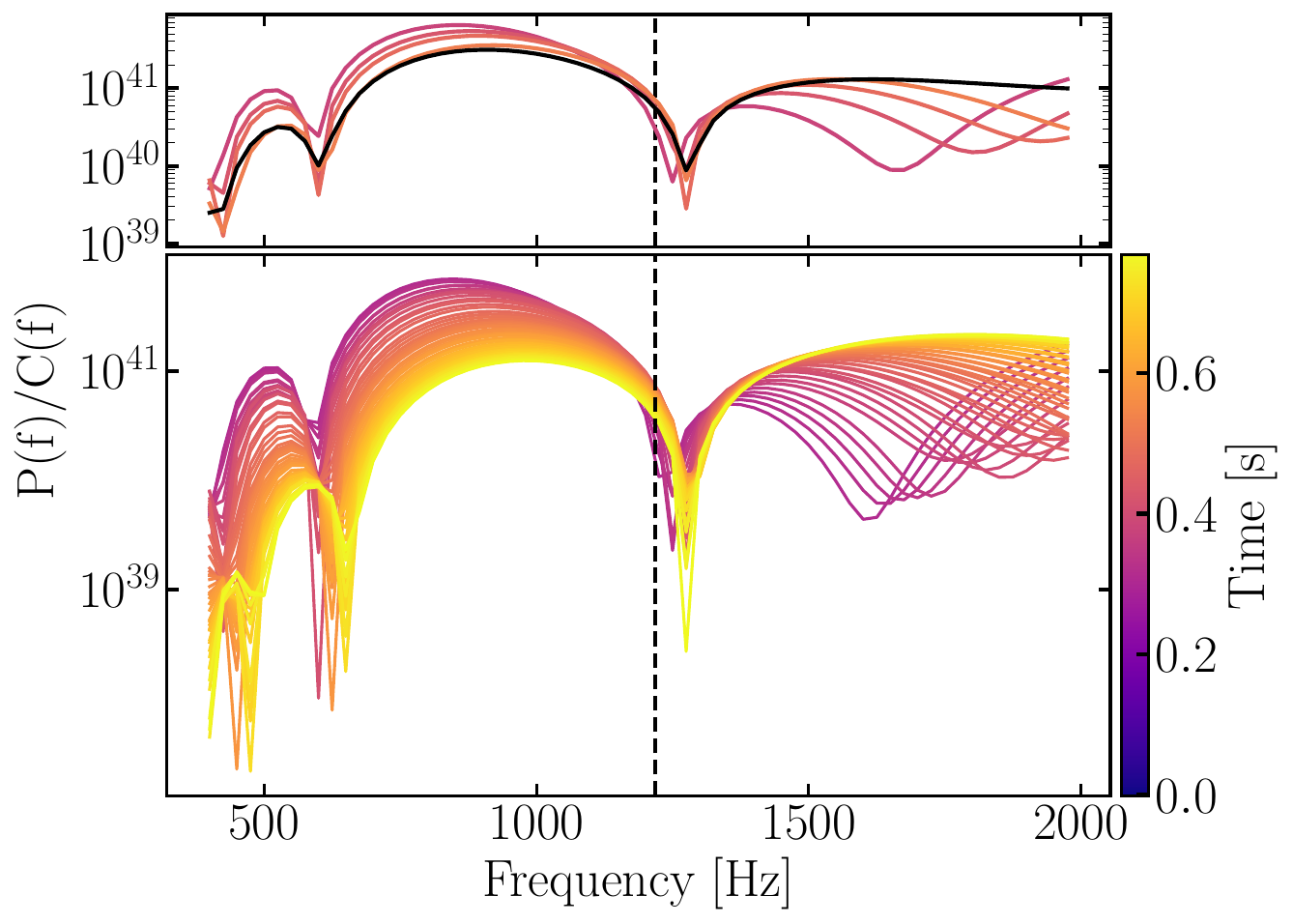} \\
    \includegraphics[width=0.49\linewidth]{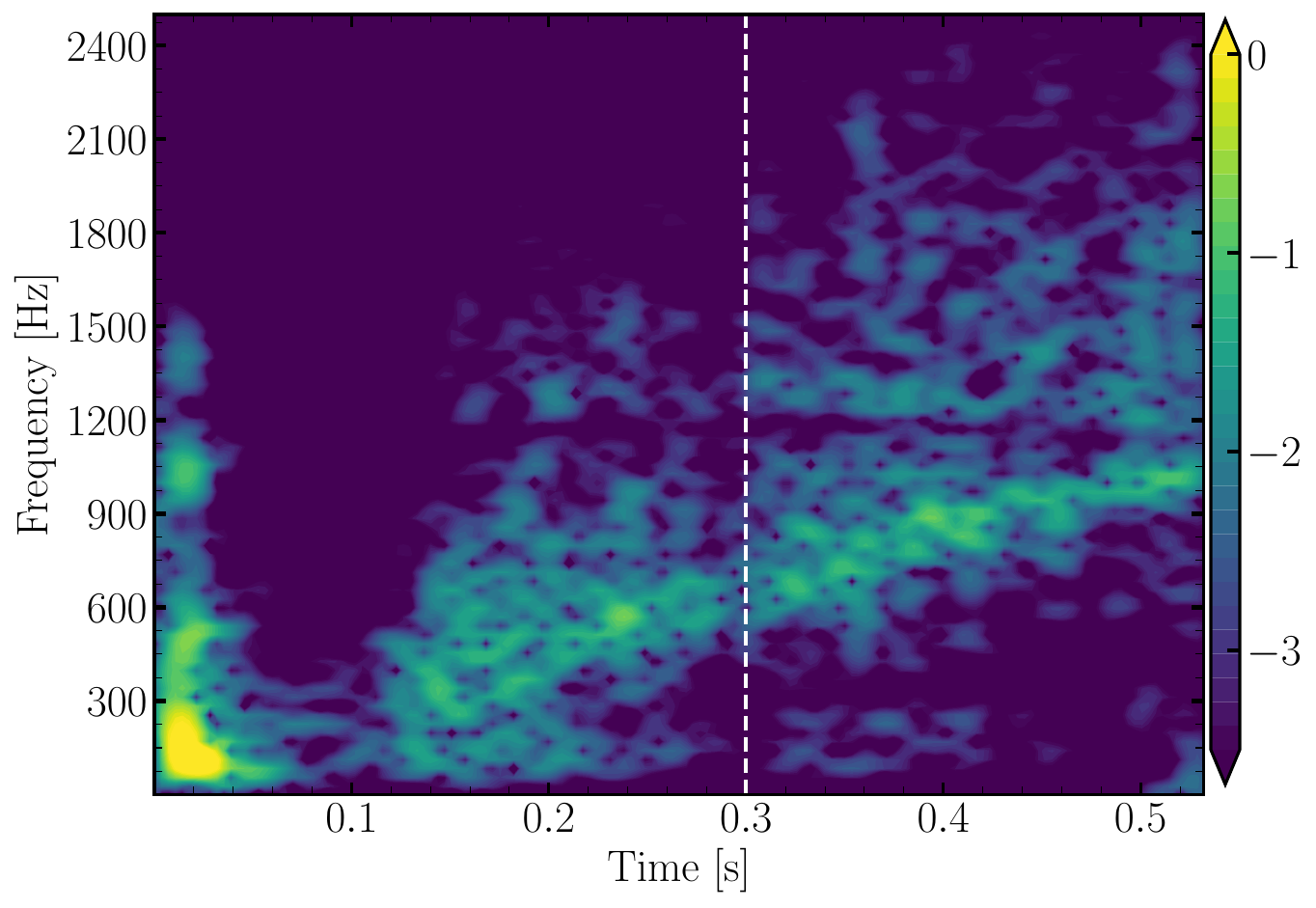}
    \includegraphics[width=0.49\linewidth]{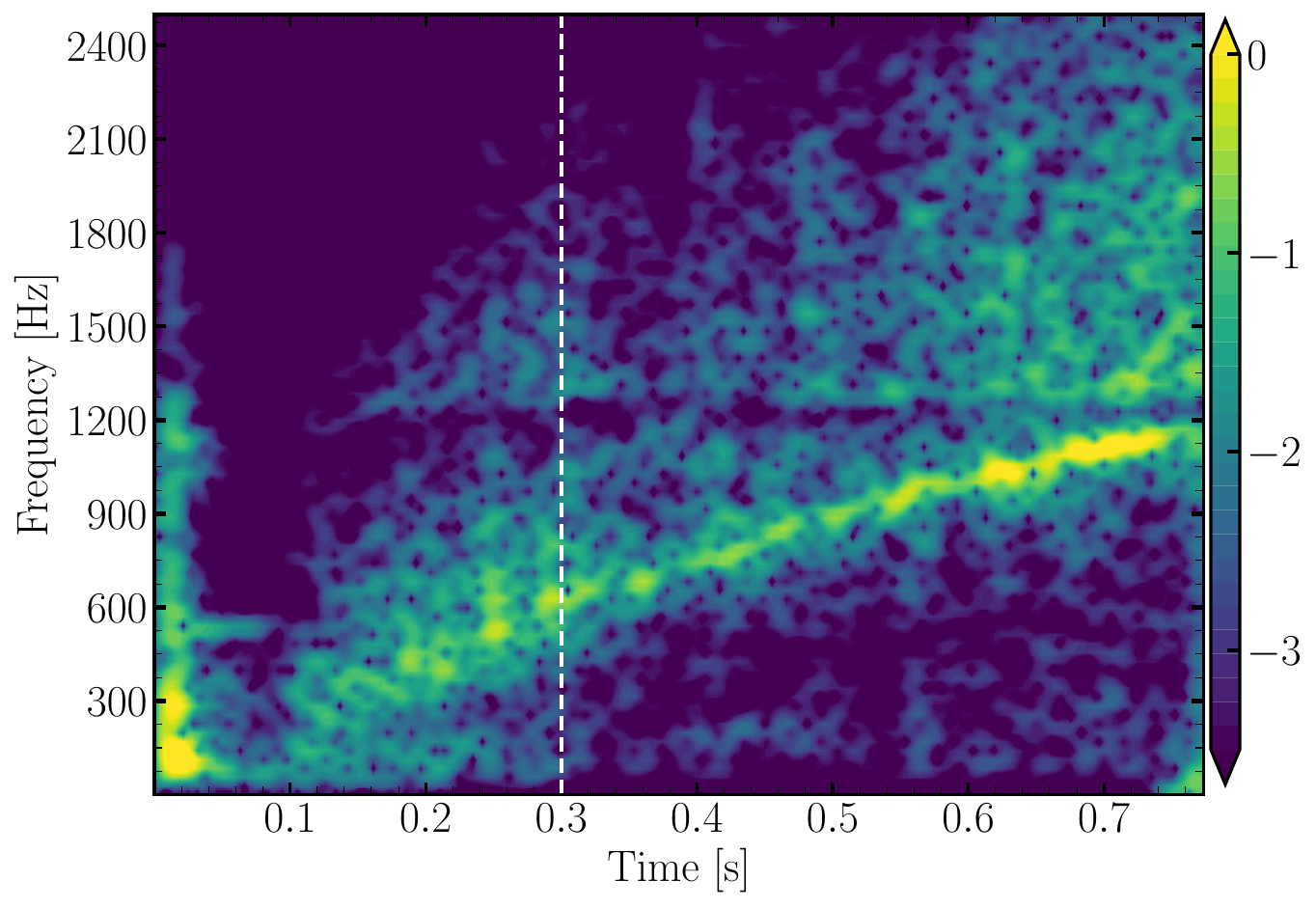}
    \caption{Power emitted as GWs by a quadrupolar perturbation (see Eq.~\ref{eq:pertpower}) between 0.1 and 0.3 s after bounce (top row) and between 0.3 s and the end of the simulation (middle row), together with the corresponding GW spectrograms (bottom row). Time is given in s after bounce. 
    The left column shows results for model s15\_95, 
    and the right column for model s20\_L45.  
    In the top and middle rows, the smaller sub-panels display the GW power averaged over 50 ms windows, and the black curves show the average GW power over the entire time interval.
    The curves in the main panels of the top and middle rows are calculated at 10 ms intervals. White vertical lines in the bottom row guide the eye to $t = 0.3$. The black lines in the two top rows indicate the average power-gap frequency extracted from our power spectral density analysis in section \ref{sec:powergap}.}
    \label{fig:spectrogramgap}
\end{figure}
Following the approach of~\cite{Zha_24}, we performed a perturbative analysis
for all models in our simulation set by solving the linear perturbation
equations over a range of frequencies and computing the corresponding GW power
spectra using Eq.~\ref{eq:pertpower}.  
The full results of this analysis were originally presented in~\cite{Li_24}.  
For most models, the resulting GW power distributions agree well with those
reported by~\cite{Zha_24}.  
In general, the perturbation analysis captures the behaviour of the power gap
for most of our models. However, a subset of models exhibits clear
differences between the perturbative predictions and the properties of the GW
signals obtained directly from the simulations.  
These discrepancies warrant further investigation, and we therefore highlight
representative examples below.

Fig.~\ref{fig:spectrogramgap} shows the spectrograms and the radiated GW power calculated according to Eq.~\ref{eq:pertpower} for models \texttt{s20\_L45} (right column) and \texttt{s15\_95} (left column). For both models,
the power gap is clearly visible in the
spectrograms from 0.15 s post bounce, even earlier in model \texttt{s20\_L45}. In contrast, the GW power spectrum predicted by the
perturbation analysis only shows a narrow gap after $\sim 300\,\mathrm{ms}$ post
bounce. The panels in the top row show Eq.~\ref{eq:pertpower} evaluated at 10 ms intervals between 0.1 s and 0.3 s post bounce. While we see clear and distinct minima in the predicted power spectra prior to 0.3 s post bounce, the frequency of the minima shifts with time. The smaller sub panels in the top row of Fig.~\ref{fig:spectrogramgap} show the same as the corresponding main panels, but averaged over 50 ms long windows or over the whole time 
window (the black lines) shown in the figure. While a local minimum for both of the two black 
curves matches the gap frequency, the average power spectrum displays a broad and shallow
valley as opposed to a clear spectral gap. The averages in 50 ms windows show more pronounced minima, but they shift around in time and span a frequency range of a few hundred 
Hertz. The situation changes 0.3 s after bounce,
see the middle row of Fig.~\ref{fig:spectrogramgap},
here we observe a strong agreement between the 
gap frequency seen in the spectrograms and the
frequency predicted by the perturbation analysis. 
Model \texttt{s20\_L45}, however, still shows some deviation between 0.3 and 0.4 s post bounce (middle-right panel).
The gap frequencies predicted by the perturbation analysis are consistently higher than the 
average power-gap frequency we estimated in section \ref{sec:powergap}, the vertical black lines in the two top rows of Fig.~\ref{fig:spectrogramgap} represent the average power-gap frequency for the two models shown in the figure. The offset is on the order of 50 Hz, which is similar to the width of the gap as seen in the spectrograms and it is possible that this discrepancy is related to a systematic offset in our method for evaluating the average frequency of the gap.

For both model \texttt{s20\_L45} and model \texttt{s15\_95}, we see clear and strong minima around 575 Hz both prior to and after 0.3 s post bounce (the two top rows of Fig.~\ref{fig:spectrogramgap}). However, these minima
do not correspond to any power gaps in the respective spectrograms (bottom row of Fig.~\ref{fig:spectrogramgap}). At late times, after 0.3 s post bounce, there is 
little GW emission below 600 Hz, it is, therefore, plausible that a gap at 575 Hz would not be visible at late times. However, both models show strong GW emission between 200 and 600 Hz before 0.3 s post bounce and a gap should in principle be visible.

The origin of the relatively poor agreement between the perturbation analysis and the spectrograms at early times is unclear. The determination of the PNS surface is more
difficult at early times and this could influence the gap frequency extracted from mode analysis. However, we do not expect it to cause a discrepancy on the level of what we
observe here.
Another possible explanation is the rapid contraction of the PNS
during this phase, which in principle violates the assumption
of a static background underlying the perturbative analysis.
However,~\cite{Murphy_25} demonstrated that hydrostatic equilibrium
is satisfied at the $\sim 10\%$ level in their simulations.
Even if hydrostatic balance is reasonably well satisfied overall,
the discrepancy between the perturbative predictions and the
spectrogram near 575 Hz may still be related to the ongoing
contraction of the PNS.
A frequency of 575 Hz corresponds to a characteristic time scale
of $\sim 1.7\,\mathrm{ms}$, compared to $\sim 0.7\,\mathrm{ms}$ for the gap near 1250 Hz.
Over longer time scales, the assumption of a
hydrostatic background may become less accurate as the PNS
continues to contract.

The idea that emission at a certain frequency cancels out when integrating over the whole PNS 
clearly requires that the emission is dominated by coherent and stable PNS oscillations. Yet, it has been demonstrated that the deceleration of downflows in the outer layers of the PNS can generate GWs at frequencies and with amplitudes matching the predicted signals~\cite{Murphy_09}. If the explanation proposed in~\cite{Zha_24} is correct it would indicate that local small-scale mass motions in isolated individual layers of the PNS are subdominant in terms of GW production. 

\subsection{Coupled modes and anti-resonances}
Systems of coupled modes, or oscillators, can exhibit a wide range of complex
behaviours beyond the avoided crossings discussed above
\cite{Aronson_90,Ida_05,Belbasi_14,Jothimurugan_16,Dolfo_18,Lockhart_18,Sarkar_19}.  
Depending on the relative coupling strength and damping, the
interaction between two modes may lead to periodic energy exchange and amplitude
beating~\cite{Bello_19}.
In such systems, the combination of oscillation modes
can give rise to interference effects
\cite{Belbasi_14,Sarkar_19,Avrutsky_13,Hayashi_23}. When this occurs, a system can
display sharp minima in its spectral power density even though all modes
remain excited~\cite{Sarkar_19}.  

One well-known manifestation of such interference is the
\textit{Fano resonance}~\cite{Fano_61},
which stems from the interaction between a narrow mode and a broad background continuum
\cite{Limonov_21,Avrutsky_13}. The Fano resonance was first proposed to explain asymmetric
line profiles in electron scattering~\cite{Fano_61} within the framework of
quantum mechanics.  Since then, Fano-type interference has been found
in a wide range of physical systems, from atomic physics and optics to plasma physics
and acoustic systems~\cite{Nozaki_13,Limonov_21}. We refer the reader to~\cite{Miroshnichenko_10} for a review and references therein for examples of
the Fano resonance.
Similar asymmetric line shapes
and antiresonances have even been demonstrated in systems of coupled classical
oscillators and analogously in electrical resonator circuits~\cite{Joe_06,Satpathy_12}.
The two coupled oscillators considered by~\cite{Joe_06} are similar to the system
described in section~\ref{subsec:crossing}, but include both a damping term
($\gamma\,\dot{x}$) and a periodic external forcing of the second oscillator
($F = A\cos{(\omega_f t)}$).  
The driven oscillator in~\cite{Joe_06} exhibits two distinct amplitude peaks,
corresponding to $\omega_{\pm}$.  
The first peak (associated with $\omega_{+}$) has a symmetric Lorentzian shape
centred near the eigenfrequency of the first oscillator.  
The second peak, however, is asymmetric and has a sharp dip where the
amplitude is nearly zero, followed by a rapid rise to a maximum near the
eigenfrequency of the second oscillator.  
This drop in amplitude occurs as a result of destructive interference between
the external driving and the oscillating system~\cite{Joe_06,Miroshnichenko_10}.

The spectral shape associated with Fano resonance can be 
written in the frequency domain as~\cite{Riffe_11,Avrutsky_13}:
\begin{equation}
P_{\mathrm{Fano}}(f) = A_0 \frac{(q+\varepsilon)^2}{1+\varepsilon^2} + B_0,
\label{eq:fano_freq}
\end{equation}
where we have assumed that the background can be described by a constant ($B_0$), the constant $A_0$ represents an overall scaling of the line, and
$\varepsilon = 2(f-f_0-\Delta)/\Gamma$. $f_0$ is the central frequency of
the anti-resonance, $\Delta$, $q$, and $\Gamma$ are parameters associated with the shape of the dip~\cite{Fano_61,Joe_06,Riffe_11,Satpathy_12}.
In complex systems, it is not trivial to prove the existence of 
the Fano resonance. A successful fit of the line shape, 
in our case the power spectral density, 
to a Fano functional form is often taken as evidence for a 
Fano-type interference mechanism~\cite{Limonov_21,Miroshnichenko_10}.
\begin{figure*}
    \includegraphics[width=0.99\linewidth]{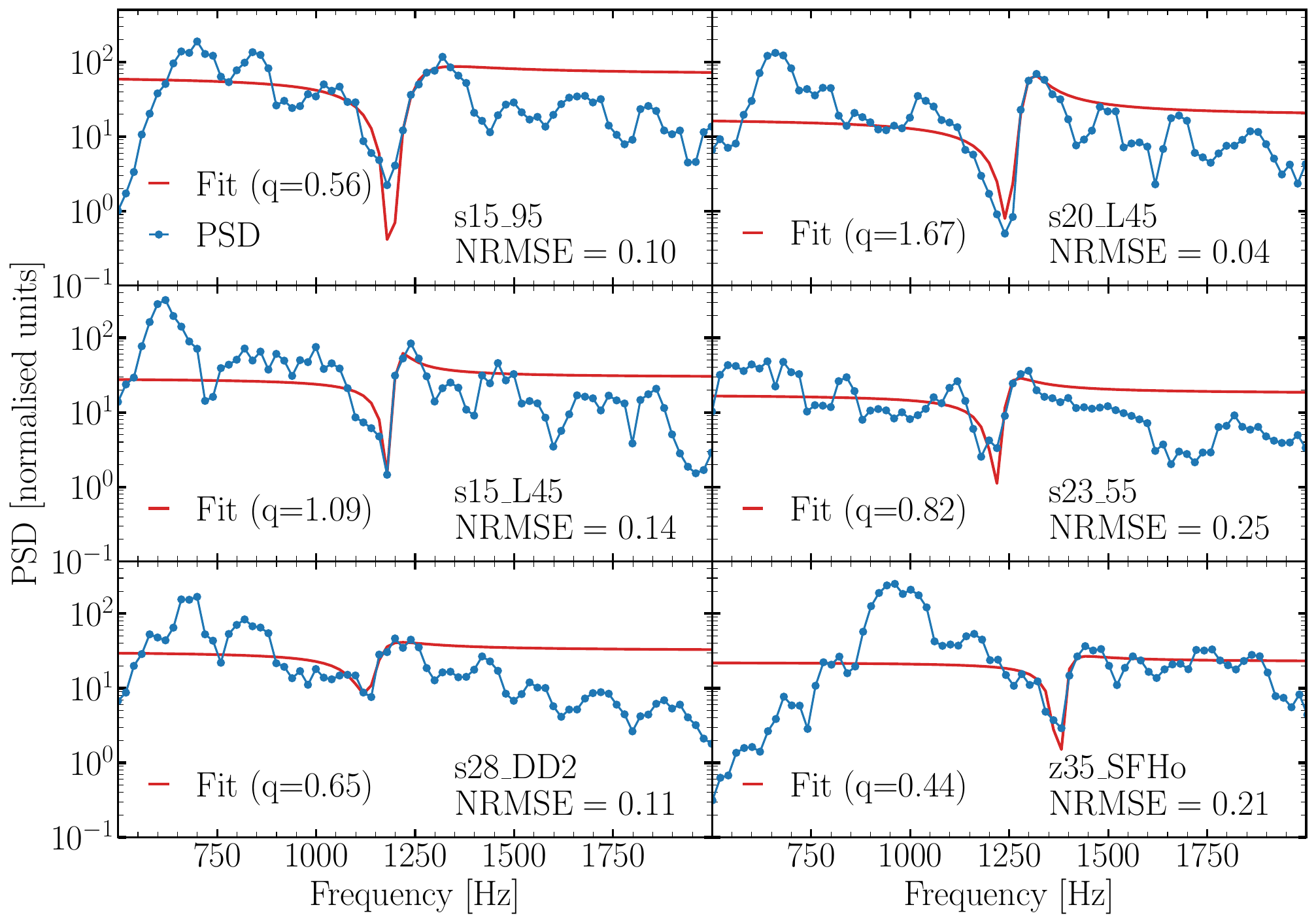}
    \caption{The PSD for six of our models (blue dots) together with the best fit to a Fano-type line profile (red lines), as a function of frequency. For each model, the PSDs are estimated based on the GW emission between 0.3 s and 0.4 s post bounce. The units of the y-axis have been normalised such that all models lie in the same range, effectively the amplitude difference between models has been normalised. The model shown in any given panel is indicated by a model label, and the $q$ parameter (see Eq.~\ref{eq:fano_freq}) is reported for each fit. {For each panel, NRMSE indicates the normalised root-mean-square error of the fit.} }
    \label{fig:fano}
\end{figure*}
To investigate whether the power gap could be interpreted as a Fano-type anti-resonance, we fit the predicted GW power spectral densities of six representative models to the line profile in Eq.~\ref{eq:fano_freq}. For each model, 
we calculate the PSD from the emission between 0.3 s and 0.4 s post bounce.
The GW signals from core-collapse supernovae can contain contributions from several noisy processes, including turbulent convection, intermittent accretion
flows, and PNS oscillations. As a result, the
short-time Fourier spectra extracted from the simulations are noisy and show stochastic
fluctuations that vary from time window to time window. We found that noise in the PSD estimates is problematic when attempting to fit the Fano profile to the data.
To address this, we use a multitaper method to obtain a PSD estimate that
better captures the underlying physical spectral content and suppresses fluctuations that arise from individual time-window realisations.
We use \texttt{scipy.signal.windows.dpss} with $\mathrm{NW}=3.5$ and
$\mathrm{Kmax}=4$ to construct the tapers for the multitaper PSD estimate.
This approach helps reduce the noise in the PSD while preserving the gap, but
we emphasise that the fitting procedure does depend on the underlying PSD estimate.
\begin{table}[t]
\centering
\caption{Fitted Fano parameters for the six models, see Eq.~\ref{eq:fano_freq}.
We do not report $A_0$ or $B_0$ as they only represent the overall scale of the problem. \label{tab:fano}}
\begin{tabular}{lcccc}
Model & $f_0$ [Hz] & $\Gamma$ [Hz]  & $q$ & $\Delta$ [Hz]   \\
\hline \hline
\texttt{s15\_95}     & 1175.49 & 136.59 & 0.56 & 51.85   \\
\texttt{s20\_L45}    & 1219.36 & 58.34  & 1.67 & 74.38   \\
\texttt{s15\_L45}    & 1169.34 & 35.41  & 1.09 & 31.21    \\
\texttt{s23\_55}     & 1140.00 & 55.56  & 0.82 & 100.00   \\
\texttt{s28\_DD2}    & 1157.70 & 83.59  & 0.65 & $-6.32$ \\
\texttt{z35\_SFHo}  & 1469.68 & 51.58  & 0.44 & $-83.58$ \\
\end{tabular}
\end{table}
The resulting fits are shown in Fig.~\ref{fig:fano} and the best fitting parameters are given in table~\ref{tab:fano}, we do not report $A_0$ or $B_0$ since they only represent an overall scale in our fitting procedure. 
In principle, the gap can be described relatively well by a Fano-type line across all models. However, the fits are far from perfect.
{To evaluate the quality of the Fano-profile fits, we use the normalised
root-mean-square error (NRMSE),
\begin{equation}
  \mathrm{NRMSE} = \frac{1}{P_{\max} - P_{\min}}
    \sqrt{\frac{1}{N}\sum_{i=1}^{N}\left(P_i - \hat{P}_i\right)^2},
  \label{eq:nrmse}
\end{equation}
where $P_i$ and $\hat{P}_i$ are the data and the fitted model, respectively, and $N$ is the number of data points. We calculate the NRMSE in a window of $\pm 100\,\mathrm{Hz}$ around the fitted
central frequency, $P_{\max} - P_{\min}$ is the range of the data in that window.
The fits yield NRMSEs of
$0.10$, $0.04$, $0.14$, $0.25$, $0.11$, and $0.21$
for \texttt{s15\_95}, \texttt{s20\_L45}, \texttt{s15\_L45}, \texttt{s23\_55},
\texttt{s28\_DD2}, and \texttt{z35\_SFHo}, respectively.
}

We see well-defined and clear gaps in all models, except in model \texttt{s28\_DD2} (bottom left panel) where the gap is shallow. The gap for model \texttt{s20\_L45} (top right panel) is
relatively wide and this is reflected in a large $\Gamma$ value, compared to the other models, see table~\ref{tab:fano}. It is also clear that the gap is not void of emission, which is particularly clear for models \texttt{s15\_95}, \texttt{s20\_L45}, and \texttt{s23\_55} where we see variation of the PSD within the gap. The Fano shape fits well to the gap for model \texttt{s28\_DD2}, even if the gap is shallow for this model.

In the canonical Fano formalism, the asymmetry parameter $q$ gives the line shape and the shift of the dip relative to the characteristic frequency of the narrowband resonant response~\cite{Fano_61, Miroshnichenko_10}. 
Large $|q|$ values produce profiles that approach a Lorentzian
shape, while $q \approx 0$ yields a symmetric anti-resonance. The sign of $q$ determines if the dip is located below or above the resonant frequency~\cite{Fano_61}, negative $q$-values correspond to a dip above the resonance and positive $q$-values to a dip below the resonance. 
As discussed by Avrutsky et al.~\cite{Avrutsky_13}, vertical offsets and scaling factors, which are often used to fit data that do not go through zero, can significantly affect the physical meaning of the fitting parameters. 
In particular, the $q$ parameter in these generalised forms is no longer a direct measure of the relative contributions of resonant and nonresonant channels, and the spectral minimum is no longer strictly located at $\epsilon = -q$.
The inclusion of background terms renders the physical interpretation of the constants more convoluted. Consequently, in our case, $q$ should be understood as a phenomenological parameter that captures the overall asymmetry of the line shape and we avoid interpreting our fit parameters in detail.

Although the Fano profile captures the qualitative structure and asymmetry of the gap, the fits should not be viewed as definitive evidence for the Fano interference producing the power gap. 
Instead, they demonstrate that an anti-resonant mechanism is \emph{compatible} with the data and may reflect the influence of more complex interference effects occurring within the PNS. 
This makes the Fano interpretation a noteworthy possibility, even if the current evidence is suggestive rather than conclusive. A Fano-type anti-resonance does not imply that the PNS is unable to oscillate at the gap frequency. It reflects a cancellation in the total GW signal. 
The quadrupolar strain receives contributions from both a broad, slowly varying background component and a narrower, more coherent mode-like component. 
The power gap could then arise due to a destructive interference between these two contributions.

A mode with a slowly evolving central frequency that tracks the power‐gap frequency is required if the gap is to be explained as an anti-resonance arising from the interaction between an eigenmode and a broadband background. Based on mode analysis, relatively stable g-mode frequencies have indeed been reported~\cite{Torres-Forne_19,sotani_20,Murphy_25}. The central frequency of $\sim 500\,\mathrm{Hz}$ identified by~\cite{Murphy_25} is likely too low to account for the observed gap, but~\cite{Torres-Forne_19} and~\cite{sotani_20} found g-modes with central frequencies near $\sim 950\,\mathrm{Hz}$, which lies much closer to the typical gap region. Morozova et al.~\cite{Morozova_18} further reported that the eigenfrequencies of the PNS core modes closely trace the power-gap frequency in their models. 
Our mode analysis identifies two eigenmodes that are associated
with the power gap.
At early times, one mode lies near the power gap,
although the correspondence is less clean and more ambiguous
than at later times.
As the PNS evolves ($t \sim 0.2$–$0.3$~s), the frequency of this mode
begins to increase.
Around the same time, it approaches a neighbouring mode
and the two undergo an avoided crossing.
Following this interaction, the frequency evolution of the second mode involved in the
crossing flattens and closely traces the power-gap evolution
at late times, while the original early-time mode
continues to increase in frequency.
It is possible that these two modes provide the stable oscillations 
required for the Fano mechanism to work.

\section{Conclusions} \label{sec:conclusions}
In this work, we analysed a large suite of axisymmetric core-collapse supernova simulations to investigate the excitation mechanism of the GW emission and physical origin of the GW power gap. We used six
progenitors and ten EOS.

We found that the GW emission in all simulations was primarily driven by turbulent accretion onto the PNS from above. Following~\cite{Radice_19}, we demonstrated a strong correlation between the total GW energy and the turbulent energy accreted by the PNS. We also showed that the haze contributed a significant fraction of the total emitted GW energy in almost all models. Although the haze is weaker than the main emission ridge at any given time, it spans a wider frequency range and extends to higher frequencies, which causes it to contribute substantially to the total energy budget. Because the haze dominates the overall emission, we investigated whether only the total energy correlated with the accreted turbulent energy or whether the two components of the signal showed similar correlations when considered separately. To test this, we examined how each component correlated with the accreted turbulent energy and with the turbulent energy inside the PNS convection zone. We isolated the two components by masking selected regions of the spectrograms prior to the Fourier analysis. Both the ridge and the haze showed strong correlations with the accreted turbulent energy, consistent with~\cite{Radice_19}, and neither component showed any correlation with the turbulent energy in the convective region of the PNS.

Out of our 60 simulations, most exhibited a clear power gap. The gap frequencies spanned a relatively narrow range across the model set, typically lying between $\sim 1000$ and
$\sim 1300\,\mathrm{Hz}$. The central gap frequency showed only a weak dependence on progenitor structure, with more compact progenitors tending to have higher gap frequencies. We also found a clear trend in which EOS that produce less compact PNS with lower central densities produced slightly lower gap frequencies.
Furthermore, a set of simulations using the SRO EOS with varying symmetry-energy slope showed that the gap frequency tended to decrease as the symmetry-energy slope increased.
A small subset of models showed no gap at all. In particular, the simulations based on the \texttt{z85} progenitor displayed no identifiable power gap, and we likewise found no gap in model \texttt{s23\_SFHx}. Several additional models showed only weak or unclear gaps that were difficult to identify by eye in the spectrograms, for example \texttt{s20\_95} and \texttt{s23\_L38}.

We investigated the correlation between various properties of the PNS and the power-gap frequency. We found that the strongest correlations were associated with quantities that characterise the inner PNS core, indicating that the gap is linked to the structure and evolution of the central region of the star. This is consistent with earlier suggestions that the mechanism responsible for the gap originates in the PNS core (e.g.,~\cite{Morozova_18,Eggenberger-Andersen_21}). In particular, we found tight relationships between the gap frequency and the central density, the central sound speed, and the Brunt–Väisälä frequency in the core. We also identified weaker correlations with the surface gravity and with the radial extent of the convective layer. 
Similar to the findings of~\cite{Morozova_18}, we identify two PNS eigenmodes that are 
associated with the power-gap frequency. In particular, at times $t \gtrsim 0.2$~s 
post bounce, the eigenmode directly above the ridge-tracking branch (in terms of frequency) traces 
the evolution of the power gap in most of our models.
{Since the power gap originates in the PNS, it is possible that a related feature could appear in the neutrino signal, or that the neutrino emission correlates with the properties of the gap. However, no such correlation has been reported in the literature. This may be because the gap appears to be connected to the inner core of the PNS, while neutrinos decouple closer to the PNS surface and therefore do not probe the same PNS regions. }

{In the event of a sufficiently nearby core-collapse supernova, the power gap would be a directly observable feature of the GW PSD, and its central frequency could therefore be used to study the properties of the PNS.}
The fact that the power gap is so tightly correlated with the properties of the PNS core means that the gap is directly linked to the properties of matter under some of the most extreme conditions throughout the universe. Measuring, for example, the sound speed in this regime could place new constraints on the EOS of hot and dense nuclear matter.
Consequently, the
power gap could provide observational insights into the properties of matter in a region of the EOS phase space that is otherwise hard to access. 
{In this context, an important question is how precisely the central frequency of a power gap could be recovered from an actual observation. The gap lies near $\sim$1~kHz, which is outside the optimal sensitivity range of current detectors. Identifying the power gap requires that the surrounding ridge and haze emission be reconstructed well enough for the  deficit in power to stand out. Its detectability, therefore, depends on how well the signal can be reconstructed in the presence of detector noise, which in turn depends on the signal-to-noise ratio, and the reconstruction method. A quantitative assessment would require injecting our signals into detector noise, reconstructing the signal, and examining PSD of the recovered signals. Such a dedicated analysis is beyond the scope of the present study, which focuses on the physical origin of the gap, but we intend to explore it in future work.}

We assessed several proposed explanations for the power gap. Avoided crossings between PNS oscillation modes are frequently invoked in the literature, but our analysis showed that it is difficult for an avoided crossing between two weakly coupled modes to generate a persistent gap in the GW emission. In such a scenario, each mode contributes power primarily near its own eigenfrequency, and the only mechanism capable of producing a suppression between the two peaks is the cross term in the Fourier spectrum arising from their coupling. However, far from the avoided crossing the coupling is weak, and the cross term becomes too small to create a pronounced or sustained minimum in the signal.

The perturbative mechanism proposed by Zha et al.~\cite{Zha_24}, in which the quadrupolar GW power integrated over the PNS vanishes at specific frequencies, reproduced the gap in many of our models and performed particularly well at late times ($\sim 0.3\,\mathrm{s}$ after bounce).
While the mechanism proposed by Zha et al.~\cite{Zha_24} provides a clear and physically motivated connection between PNS structure and the appearance of a spectral minimum, we found considerably worse agreement between the predicted gap frequency and the gap frequency observed in our models during the early post-bounce phase. During this phase, the perturbation analysis predicted a significant drift in the location of the power gap, but the power gap observed in the simulations remained relatively stable during this time.
In addition, the perturbation analysis predicted an extra minimum near 550 Hz in several models, but we found no evidence for a corresponding power gap in that frequency range. 

Finally, we investigated whether the gap could arise from an anti-resonance between different contributions to the GW signal. To test this idea, we fitted a Fano-type interference profile~\cite{Fano_61} to the GW PSDs and obtained reasonably good fits. We did not interpret this as conclusive evidence for a Fano mechanism, but rather as a demonstration that interference between a coherent oscillation and a broadband background can, in principle, generate a narrow suppression similar to the observed gap.

A complete explanation of the power gap likely requires
detailed analysis of the dynamics of non-linearly coupled modes in the presence of turbulent forcing. 
Finally, we caution that all results presented in this work are
based on axisymmetric (two-dimensional) simulations.
The dynamics of core-collapse supernovae are inherently
three-dimensional. Andresen et al.~\cite{Andresen_17}
connected differences in the GW excitation and emission, between two and three dimensional
simulations, to fundamental differences in the hydrodynamics of axisymmetric simulations and
full three-dimensional simulations. The coupling between the turbulent flow
inside and outside of the PNS to mode excitation and, in turn, GW emission can differ
in two versus three dimensions.
It is, therefore, possible that several aspects of the power-gap and its
correlations to the underlying physical properties of the simulations could change in three dimensions.
{However, the properties of the gap is set by the structure of the PNS which is less sensitive to the dimensionality of the simulations. On the other hand, full general relativity could modify the PNS structure and change the PNS mode frequencies, which would at the very least lead to a shift in the results reported here. To date, the power gap has not been reported in any fully relativistic simulation.}
Future investigations will be required to assess
the robustness of our findings in fully three-dimensional
simulations.
{A natural extension of our work would be to carry out a direct comparison between the properties of the power gap in two-, and three-dimensional simulations of the same progenitor. However, the power-gap has been seen in three-dimensional simulations (see for example Vartanyan et al. \cite{Vartanyan_23}.}

\ack{
We are grateful to Mark Riffe for answering questions regarding the Fano mechanism. We thank André da Silva Schneider for providing the SRO EOS tables.
The computations were enabled by resources provided by the National Academic Infrastructure for Supercomputing in Sweden (NAISS) at NSC partially funded by the Swedish Research Council through grant agreements no. 2022-06725 and no. 2018-05973. 
\textbf{{Software}:} FLASH~\cite{Fryxell_2000}, NuLib~\cite{OConnor_15}, Matplotlib~\cite{Hunter07}, NumPy~\cite{numpy}, SciPy~\cite{2020SciPy-NMeth}, yt~\cite{Turk11}.
}

\funding{This work is supported by the Swedish Research Council (Project No. 2020-00452). S. Z. received support from the Yunnan Revitalization Talent Support Program--Young Talent project, and the Yunnan Fundamental Research Projects (grant No. 202501AS070078).}

%\roles{}

\data{The GW signals from all 60 models presented in this study are available at \url{https://zenodo.org/records/20607405}. Additional derived data presented in this work, analysis tools, and the underlying simulation can be provided upon reasonable request.}

\suppdata{}

\bibliographystyle{iopart-num}
\bibliography{sources}

\end{document}